\documentclass{article}

\usepackage{amsmath,amsfonts,graphicx,amsthm}
\usepackage[margin=1.5in]{geometry}

\usepackage{geometry}
\usepackage{color}
\usepackage{amsfonts}
\usepackage{graphics}
\usepackage{amsmath}
\usepackage{graphicx}
\usepackage{multirow}
\usepackage{pifont}
\usepackage{float}
\usepackage[authoryear]{natbib}
 \bibpunct{(}{)}{;}{a}{,}{,}

\usepackage{enumitem}

\newtheorem{theorem}{Theorem}
\newtheorem{corollary}{Corollary}
\newtheorem{lemma}{Lemma}

\newcommand\independent{\protect\mathpalette{\protect\independenT}{\perp}}
\def\independenT#1#2{\mathrel{\rlap{$#1#2$}\mkern2mu{#1#2}}}

\usepackage{bbm}
\newcommand{\ind}{\mathbbm{1}}

\title{Inference for natural mediation effects under case-cohort sampling with applications in identifying COVID-19 vaccine correlates of protection}

\author{D BENKESER, I DIAZ, J RAN}

\begin{document}

\maketitle

\begin{abstract}
Combating the SARS-CoV2 pandemic will require the fast development of effective preventive vaccines. Regulatory agencies may open accelerated approval pathways for vaccines if an immunological marker can be established as a mediator of a vaccine's protection. A rich source of information for identifying such correlates are  large-scale efficacy trials of COVID-19 vaccines, where immune responses are measured subject to a case-cohort sampling design. We propose two approaches to estimation of mediation parameters in the context of case-cohort sampling designs. We establish the theoretical large-sample efficiency of our proposed estimators and evaluate them in a realistic simulation to understand whether they can be employed in the analysis of COVID-19 vaccine efficacy trials.
\end{abstract}


\section{Introduction}

The best hope for stemming the SARS-CoV2 pandemic is to develop a safe and effective preventive vaccine and to distribute the vaccine widely \citep{corey2020strategic}. Several  vaccines have already been demonstrated to exhibit high efficacy and have been granted emergency use authorizations around the world \citep{baden2021efficacy, polack2020safety}. However, supply issues for COVID-19 vaccines remain a pressing concern. Thus, it is of critical importance to continue to bring new effective vaccines to the market. 

An important means of accelerating the vaccine development process is to establish immune correlates of vaccine efficacy. Immune correlates are immunogenicity assays that are predictive of the vaccine's effect on infection with SARS-CoV2 and/or COVID-19 disease \citep{qin2007framework}. These immune responses are typically measured shortly after receipt of the final dose of a vaccine. For SARS-CoV2, scientists are keenly interested in antibody responses to the vaccine insert SARS-CoV-2 proteins, including binding, pseudovirus neutralizing, and live virus neutralizing antibodies. Discovery and validation of a strong immune correlate would have an immense impact on SARS-CoV2 vaccine development, as it could provide a valid surrogate endpoint and thereby open an expedited regulatory pathway for licensure of vaccine products. Currently, in order to adequately power a Phase III licensure trial in the U.S., studies are enrolling between 30,000 and 60,000 individuals and following for up to two years. If a licensure pathway were opened based on an immune correlate, licensure trials could be designed to yield results in weeks rather than years. 

Because of the importance of immune correlates to the success of a vaccine development program, sub-studies of ongoing Phase III randomized trials are designed specifically to examine immune correlates in a harmonized fashion \citep{ows_sampling_plan}. There is a wide body of literature available on statistical approaches for establishing immune correlates. These methods broadly fall into two classes: correlates of risk and correlates of protection (also known as, respectively, non-mechanistic and mechanistic correlates of protection). The former approaches seek to evaluate relative and absolute risks of infection and disease across levels of an immune response. These methods can be framed as a standard supervised learning problem, where one is predicting infection/disease status based on one or several sets of participant information. Mechanistic correlates of protection, on the other hand, seek to establish a causal relationship between vaccination, immune response and outcome. There is a rich literature available on statistical approaches based on principal stratification \citep{gilbert2008evaluating,joffe2009related,li2010bayesian,wolfson2010statistical}. An alternative approach is to evaluate immune responses as {\it mediators} of the vaccine's effect \citep{cowling2019influenza}.

Mediation is a very deep literature with contributions coming from across many diverse fields over several decades. \citet{vanderweele2016mediation} and \citet{nguyen2020clarifying} provide excellent overviews of current research in the area. In this work, we focus on {\it natural} mediation parameters \citep{robins1992identifiability,pearl2001direct}. These parameters decompose a vaccine's effect into {\it direct} pathways (i.e., pathways not involving the measured immune responses) and {\it indirect} pathways (i.e., pathways through the measured immune responses). Natural indirect effects involve a comparison of two counterfactual risks: (i) the risk of infection or disease under a hypothetical intervention wherein individuals receive the vaccine, but rather than allowing the vaccine to determine the level of immune response, we fix their immune responses to the values they would naturally assume under placebo; (ii) the risk of infection or disease when individuals receive the vaccine and their immune responses are allowed to naturally respond to the vaccine. Comparison of these risks yields one causal effect that quantifies the impact of the immune response on risk of infection and disease.

Identification and estimation of natural mediation effects has been extensively studied (among others, \citet{petersen2006estimation,van2008direct,imai2010identification,tchetgen2012semiparametric,zheng2012targeted}). However, none of these approaches is well-suited for use in the sampling frame of COVID-19 vaccine efficacy trials, where immune responses are measured using {\it case-cohort} sampling \citep{breslow2013using}. In this approach, all Phase III trial participants have blood drawn shortly after receipt of the final vaccination. However, immune responses are only measured in a stratified random sub-sample (i.e., cohort) of individuals, with inclusion probabilities dependent on participants' age, race/ethnicity, and SARS-CoV2 baseline serostatus. In order to adequately power correlates analysis,  data are augmented to include immune responses from all individuals who eventually become infected with SARS-CoV2 (i.e., cases). Thus, the observed data represent a biased sub-sample of the trial population.

In this work, we describe methods for estimation and inference on natural mediation effects in this context. 
Our work builds on past work on multiply robust estimation \citep{tchetgen2012semiparametric,zheng2012targeted}. We combine these approaches with theory of two-phase sampling \citep{rose2011targeted} to establish the nonparametric efficiency bound for regular, asymptotically Normal estimators in this problem. We establish the efficient influence function of these parameters in a nonparametric model, which naturally suggests one-step estimators of the effects of interest. Using these results we propose two multiply robust estimators. One is derived using a direct application of the strategy suggested by \citet{rose2011targeted}. The second is novel in that it relies on a new representation of the efficient influence function. For both estimators, we establish regularity conditions under which they converge weakly to Normal random variables. Furthermore, we study and compare their performance in finite samples using simulations targeted towards the design of the randomized COVID-19 vaccine efficacy trials. 

\section{Background}
\subsection{Natural mediation effects}

Suppose that we have access to a randomized trial that generates independent observations of the vector $X = (W, A, S, C, CY) \sim P_X$, where $W$ is a vector of baseline covariates, $A$ is the randomized vaccine assignment ($A=0$ denotes randomization to placebo, $A=1$ to vaccine), $S$ is a vector of immune responses, $C$ is an indicator of full follow-up for a participant (i.e., that the participant is {\it not} lost-to-followup), and $Y$ is the study endpoint (e.g., infection with SARS-CoV2 or clinical COVID-19 disease). Here, we represent the data in such a way that the outcome $CY$ is set to zero is $C = 0$, though this is arbitrary and does not affect our subsequent developments. Note that this trial is hypothetical, because in reality we will not have observed the immune responses $S$ for all trial participants.

We assume a model for $P_X$ that makes certain positivity assumptions on the conditional distributions of $A$, $S$ and $C$. Specifically, for $a_0 = 0, 1$, we define $g_{A \mid W}(a_0 \mid W) := P_X(A = a_0 \mid W)$ and assume a positivity condition, $P_X\{ \delta_A < g_{A \mid W}(a_0 \mid W) < 1 - \delta_A \} = 1$ for some $\delta_A > 0$. With respect to censoring, we assume that $C$ does not depend on $S$, which seems reasonable in our context since individuals are blinded to the value of their immune response measurements during the trial. The methods below could easily be adapted to handle the case where censoring depends on $S$. Moreover, we define $g_C(1 \mid a_0, w) := P_X(C = 1 \mid A = a_0, W = w)$ and assume a positivity condition $P_X\{g_C(1 \mid a_0, W) > \delta_C \} = 1$ for some $\delta_C > 0$. Finally, we make a positivity assumption on the conditional immune response distribution. For $a_0 = 0,1$, we denote by $q_{S \mid a_0, W}(\cdot \mid W)$ the conditional density of $S$ given $A = a_0, W$. We assume for $a_1 = 0, 1$ and $a_2 = 1 - a_1$, $P_X\left[\ \{q_{S \mid a_2, W}(S \mid W)\ / \ q_{S \mid a_1, W}(S \mid W) ] < \infty \right\} = 1,$ which is an assumption of common support of the $W$-conditional immune response distributions between vaccinated and placebo-recipients. It is possible that this assumption could be violated in the context of COVID-19 vaccine trials for certain vaccines and particular immune responses. In particular, for a highly immunogenic vaccine, it may be that all participants have a positive immune response. In this case, there would be positive mass at zero in the immune response distributions amongst SARS-CoV2 seronegative placebo recipients, while amongst SARS-CoV2 seronegative vaccine recipients there would always be a positive immune response. In such cases, one possible path forward is to re-frame the mediation question, for example, by defining $S$ as a binary fold-rise in antibody response. 

Beyond these positivity conditions, our model makes no assumption on the distribution $P_X$. However, our results apply to settings where additional assumptions are placed on $g_{A \mid W}$ and $g_C$, including the possibility that these quantities are known exactly. 

We define $Y(a, s)$ as the value $Y$ would assume under a hypothetical intervention that sets $A = a$, $S = s$, and $C = 1$, though our notation suppresses this latter intervention for simplicity. Thus, we can define a counterfactual observation $X^* = (W, S(a_1), Y(a_2,s) : (a_1, a_2) \in \{0,1\}^2, s \in \mathcal{S}) ) \sim P^*_X$ for some set $\mathcal{S}$. 

We use $E_X^*\{f(X^*)\} = \int f(x) dP^*_X(x)$ to denote the expectation of any $P^*_X$-measurable function $f$ under $P^*_X$. The effect of the vaccine can be quantified in terms of a ratio of counterfactual risks, $E_X^*\{Y(1,S(1))\}/E_X^*\{Y(0,S(0))\}$. Notice that this is a contrast of average counterfactual outcomes generated under two distinct interventions. The first intervention (in the numerator) randomizes a participant to receive vaccine and then sets their immune response to $S(1)$, that is, the value their immune response would naturally assume under vaccine. The second (in the denominator), similarly randomizes to placebo and sets immune responses to their natural value under placebo. This effect can be decomposed as follows: \begin{equation*}
    \frac{E_X^*\{Y(1,S(1))\}}{E_X^*\{Y(0,S(0))\}} = \frac{E_X^*\{Y(1,S(1))\}}{E_X^*\{Y(1,S(0))\}} \  \frac{E_X^*\{Y(1,S(0))\}}{E_X^*\{Y(0,S(0))\}} \ ,
\end{equation*}
where the first term is a natural indirect effect and the latter a natural direct effect. The former compares counterfactual risks under interventions that only differ in how they assign $S$ and thus describes the impact of modulating $S$ on the protective efficacy of the vaccine. The latter compares interventions that only differ in the vaccination status, thereby evaluating the impact of the vaccine that is not influenced by $S$. For brevity, we hence discuss identification and estimation of $\psi^*(a_1, a_2) := E_X^*\{Y(a_1,S(a_2))\}$, for $(a_1, a_2) \in \{0,1\}^2$, noting that each of these effects can be written in terms of these quantities. 

Under certain causal assumptions, one can show that $\psi^*(a_1,a_2) = \psi(a_1, a_2)$, where, using $E_X$ to denote expectation under $P_X$, \begin{equation} \label{id_form}
 \psi(a_1, a_2) := E_X[ E_X \{E_X(CY \mid A = a_1, C = 1, S, W) \mid A = a_2, W\} ] \ .
\end{equation}
The form of these assumptions has been the subject of debate; see e.g., \citet{zheng2012targeted}. We omit such a discussion here and focus instead on the statistical estimation problem of the parameter (\ref{id_form}).

\subsection{Review of asymptotic linearity and influence functions}

An estimator $\hat{\psi}(a_1, a_2)$ of $\psi(a_1, a_2)$ is said to be {\it asymptotically linear} if $\hat{\psi}(a_1, a_2) - \psi(a_1, a_2) = n^{-1} \sum_{i=1}^n D(P_X)(X_i) + o_{\text{p}}(n^{-1/2})$ for some function $D(P_X)$ such that $E_X\{D(P_X)(X)\} = 0$ and $E_X\{D^2(P_X)(X)\} < \infty$. The function $D(P_X)$ is called the {\it influence function} of $\hat{\psi}(a_1, a_2)$ and is a key object for (i) characterizing the asymptotic sampling distribution of the estimator; (ii) establishing an efficiency bound for regular estimators of $\psi(a_1, a_2)$; (iii) describing robustness properties of efficient estimators; (iv) constructing efficient estimators. 

With respect to the sampling distribution of the estimator, we note that asymptotically linear estimators are particularly convenient for asymptotic analysis, as their large sample behavior is described by the weak law of large numbers and the central limit theorem. The latter implies that $n^{1/2}\{\hat{\psi}(a_1, a_2) - \psi(a_1, a_2)\}$ converges weakly to a random variable with a mean-zero Normal distribution with variance $E_X\{D^2(P_X)(X)\}$. Influence functions are also useful for establishing local efficiency of estimators. Because the asymptotic variance of an asymptotically linear estimator equals the variance of the its influence function, we can often describe an {\it efficient} influence function, that is, the influence function that has the smallest variability amongst all influence functions of regular, asymptotically linear estimators. An estimator with influence function equal to the efficient influence function is said to be asymptotically efficient. The influence function can be used to describe which combinations of parameters of $P_X$ must be consistently estimated in order to achieve a consistent estimator of $\psi(a_1,a_2)$. For many parameters, one can obtain a consistent estimator of the parameter of interest with two or more distinct combinations. We describe such parameters and estimators as doubly- or multiply-robust. 


Influence functions are also used to construct asymptotically linear and efficient estimators via the one-step estimation framework. Suppose we have an estimate of $P_X$, say $\hat{P}_X$, that we use to construct a {\it plug-in} estimate of $\psi(a_1, a_2)$, say $
\hat{\psi}(a_1, a_2) := \hat{E}_X[ \hat{E}_X \{\hat{E}_X(CY \mid A = a_1, C = 1, S, W) \mid A = a_2, W\} ]$,
where $\hat{E}_X$ denotes expectation under $\hat{P}_X$. A one-step estimator can be constructed as $\hat{\psi}^+(a_1, a_2) := \hat{\psi}(a_1, a_2) + n^{-1}\sum_{i=1}^n D(\hat{P}_{X})(X_i)$. Under regularity conditions on $\hat{P}_X$, $\hat{\psi}^+(a_1, a_2)$ can be shown to be asymptotically linear and efficient. In practice, as we will soon see, it is often not necessary to estimate the entire distribution $P_X$, but rather only key parameters of the distribution, which we refer to as {\it nuisance parameters}. 

The efficient influence function for $\psi(a_1, a_2)$ is provided in \citet{tchetgen2012semiparametric}. To write down its form, we first require notation to describe key nuisance parameters. We define $g_{A \mid W, S}(a_0 \mid W, S) := P_X(A = a_0 \mid W, S)$, the conditional probability of randomization status given $W$ and $S$, as well as $\bar{Q}_{Y}(W,S) := E_X(Y \mid A = a_1, W, S)$, the conditional probability of the outcome. Subsequently, we can define $\bar{Q}_{\bar{Q}_Y(W,S)}(W) := E_X\{\bar{Q}_{Y}(W, S) \mid A = a_2, W \}$ as the conditional mean of $\bar{Q}_Y(W,S)$ given $A = a_2, W$. Finally, we use $Q_W(w) := P_X(W \le w)$ to denote the cumulative distribution function of $W$. Using these definitions, we can write the nonparametric efficient influence function for $\psi(a_1, a_2)$ evaluated on an observation $x$ as \begin{equation}\label{fulldata_eif} \begin{aligned}
D(P_X)(x) &:= \frac{\ind(a = a_1, c = 1)}{g_{A \mid W}(a_2 \mid w) g_{C}(1 \mid a_1, w)} \frac{g_{A \mid W,S}(a_2 \mid w, s)}{g_{A \mid W,S}(a_1 \mid w, s)} \{ y - \bar{Q}_{Y}(w,s) \} \\
&\hspace{0.4in} + \frac{\ind(a = a_2)}{g_{A \mid W}(a_2 \mid w)} \{\bar{Q}_{Y}(w,s) - \bar{Q}_{\bar{Q}_Y(W,S)}(w) \} \\
&\hspace{0.6in} + \bar{Q}_{\bar{Q}_Y(W,S)}(w) - \int \bar{Q}_{\bar{Q}_Y(W,S)}(w) \ dQ_W(w) \ . 
\end{aligned}
\end{equation}
The efficient influence function is multiply robust in the sense that it is a valid estimating equation for $\psi(a_1, a_2)$ if any of the following conditions holds: (i) $\bar{Q}_Y$ and $\bar{Q}_{\bar{Q}_Y(W,S)}$ are correct; (ii) $g_{A \mid W, S}$, $g_{A \mid W}$, and $g_C(1 \mid a_1, \cdot)$ are correct; or (iii) $\bar{Q}_Y$, $g_{A \mid W}$, and $g_C(1 \mid a_1, \cdot)$ are correct. Here we use ``correct'' to mean equal to the true regression implied by $P_X$. When it comes to estimation, this robustness will imply that we only need to consistently estimate certain combinations of these nuisance parameters to obtain a consistent estimate of $\psi(a_1, a_2)$. 

\subsection{Two-phase sampling designs}

In the context of COVID-19 vaccines, $S$ is measured subject to case-cohort sampling. The observed data can thus be represented as $n$ independent copies of $O = (W, A, R, RS, C, CY) \sim P$, where $R$ is an indicator of having $S$ measured. As with $Y$, we arbitrarily set $S = 0$ if immune responses are not measured, though this does not impact subsequent developments. We denote by $g_{R}(1 \mid W, A, C, CY) = P(R = 1 \mid W, A, C, CY)$ the probability of having immune responses measured. Recall that, because all cases are sampled by design, $g_{R}(1 \mid w, a, 1, 1) = 1$ for all $w$ and $a$, but otherwise will equal to the probability of being randomly selected given covariates and vaccine assignment. The model for $P$ can be defined in terms of the model described above for $P_X$ and a model for the sampling mechanism. The only assumption we make on the sampling mechanism is that $P\{ g_{R}(1 \mid W, A, C, CY) > 0 \} = 1$, which is satisfied by design in COVID-19 vaccine trials.

Rose and van der Laan (2011) provide a convenient characterization of the efficient influence function for pathwise differentiable parameters in two-phase sampling settings. Since $\psi(a_1, a_2)$ is indeed pathwise differentiable and case-cohort sampling is a special case of general two-phase sampling, we can apply these results directly. Their results imply that the efficient influence function of $\psi(a_1,a_2)$ in a nonparametric observed data model for $P$, when evaluated on a typical observation $o$ can be written \begin{equation} \label{obsdata_eif}\begin{aligned}
D(P)(o) &:= \frac{r}{g_{R}(w, a, c, cy)} D(P_X)(o) \\
&\hspace{0.1in} + \left\{ 1 - \frac{r}{g_{R}(w, a, c, cy)}\right\} E\{ D(P_X)(O) \mid R = 1, W = w, A = a, C = c, CY = cy \} \ .
\end{aligned}
\end{equation}
Above, we only need evaluate $D(P_X)$ on observations $O$ such that $R = 1$ and when $R = 1$ $O = X$ and thus we are able to evaluate $D(P_X)(O)$. Because $D(P_X)$ depends on parameters of the full data distribution, in the next section we establish identifiability of these quantities based on the observed data distribution. 

The influence function (\ref{obsdata_eif}) is doubly robust in the sense that it is a valid estimating equation for $\psi(a_1, a_2)$ if $D(P_X)$ is a valid estimating function in the full data model and {\it either} $g_R$ or $E\{ D(P_X)(O) \mid R = 1, W = w, A = a, C = c, CY = cy \}$ corresponds to the true value of the regressions implied by $P$.

\section{Methods}


\subsection{A classic one-step estimator}

A one-step estimator can be constructed directly from the results of Rose and van der Laan (2011). Recall that one-step estimators involve two ingredients: (i) a plug-in estimator of the target parameter and (ii) the empirical average of its efficient influence function at the estimated nuisance parameters evaluated on the observed data. 

With respect to (i), we note that a plug-in estimator of $\psi(a_1, a_2)$ can be obtained via inverse probability weighted sequential regression. First, we require an estimate $g_{n,R}$ of $g_R$, if it is unknown. This could be obtained using regression of the binary outcome $R$ onto $W, A, C,$ and $CY$. The estimated sampling probabilities are then used to estimate $\bar{Q}_{Y}$. Here, we regress the outcome $Y$ onto covariates $W$ and $S$ amongst the subset of data with $R = 1$ and $A = a_1$, while including inverse probability weights $R / g_{n,R}(1 \mid W, A, C, CY)$ into the learning procedure. For example, logistic regression working models could be used  with regression parameters estimated by maximizing an inverse weighted log-likelihood. More flexible learning approaches could also be adopted. Irrespective of the learning approach chosen, we let $\bar{Q}_{n, Y}$ denote the estimated regression function. The second step in the procedure involves evaluating $\bar{Q}_{n, Y}(W_i, S_i)$ for $i$ such that $R_i = 1$. This prediction serves as the outcome in a regression onto covariates $W$ amongst the subset of data with $R = 1, A = a_2$, again including inverse probability weights. Let $\bar{Q}_{n, \bar{Q}_Y(W,S)}$ denote the estimated regression. A plug-in estimate is obtained as $\psi_{n,1}(a_1, a_2) = n^{-1} \sum_{i=1}^n \bar{Q}_{n, \bar{Q}_Y(W,S)}(W_i)$. 

Inspection of (\ref{obsdata_eif}) reveals that it involves both the full data efficient influence function (\ref{fulldata_eif}) (in the first line), as well as an additional nuisance parameter $\bar{Q}_{D(P_X)}(W, A, C, CY) := E\{ D(P_X)(O) \mid R = 1, W, A, C, CY\}$ (second line). To evaluate the full data efficient influence function, we will need estimates of $g_{A \mid W}$, $g_{C}$, and $g_{A \mid W, S}$. The randomization probability $g_{A \mid W}$ is known our example, but in general could be estimated. To maintain generality, we denote by $g_{n,A \mid W}$ the estimated regression function. An estimate $g_{n,C}$ of the censoring probability $g_{C}$ could be obtained via regression of the binary outcome $C$ onto $W$ amongst observations with $A = a_1$. Importantly these two regression, {\it do not} require inverse probability weights, since the distribution of $A, W,$ and $C$ are observed for everyone. On the other hand, an estimate $g_{n, A \mid W, S}$ of $g_{A \mid W, S}$ would need to involve inverse probability weights, for example as in a weighted regression of the binary outcome $A$ onto $W$ and $S$. 

Once all of the above regressions have been estimated, we can replace the true nuisance parameters in (\ref{fulldata_eif}) with their estimated counterparts and evaluate the expression for all $O_i$ such that $R_i = 1$,\begin{equation} \label{eval_eif_at_estimates} \begin{aligned}
    \hat{D}_X(O_i) &:= \frac{\ind(A_i = a_1, C_i = 1)}{g_{n, A \mid W}(a_2 \mid W_i) g_{n, C}(1 \mid a_1, W_i)} \frac{g_{n, A \mid W,S}(a_2 \mid W_i, S_i)}{g_{n, A \mid W,S}(a_1 \mid W_i, S_i)} \{ Y_i - \bar{Q}_{n, Y}(W_i, S_i) \} \\
&\hspace{0.4in} + \frac{\ind(A_i = a_2)}{g_{n, A \mid W}(a_2 \mid W_i)} \{\bar{Q}_{n, Y}(W_i, S_i) - \bar{Q}_{n, \bar{Q}_Y(W,S)}(W_i) \} \\
&\hspace{0.6in} + \bar{Q}_{n, \bar{Q}_Y(W,S)}(W_i) - \psi_{n,1}(a_1, a_2) \ . 
\end{aligned}
\end{equation}
Next, we obtain an estimate $\bar{Q}_{n,D(P_X)}$ of the additional nuisance parameter $\bar{Q}_{D(P_X)}$ by regressing the pseudo-outcome $\hat{D}_X(O_i)$ onto $W_i, A_i,$ $C_i, C_iY_i$ amongst observations with $R_i = 1$. The estimated efficient influence function is \begin{align*}
\hat{D}_1(O_i) &:= \left[ \frac{R_i}{g_{n,R}(1 \mid W_i, A_i, C_i, C_iY_i)} \hat{D}_X(O_i) \right. \\
 &\hspace{1.2in} + \left\{ 1 -  \frac{R_i}{g_{n,R}(1 \mid W_i, A_i, C_i, C_iY_i)} \right\} \bar{Q}_{n,D(P_X)}(W_i, A_i, C_i, C_i Y_i) \bigg] \ . 
\end{align*}
The one-step estimator is $\psi_{n,1}^+(a_1, a_2) := \psi_{n,1}(a_1, a_2) + n^{-1} \sum_{i=1}^n \hat{D}_1(O_i)$.

We have the following theorem describing the behavior of the one-step estimator. We write the squared $L^2(P)$-norm of a square-integrable function $f$ as $|| f ||^2_P := \int f(o)^2 dP(o)$. Let $\bar{Q}_{D(\hat{P}_X)}(W, A, C, CY) := E\{ \hat{D}_1(O) \mid W, A, C, CY \}$ and $\bar{Q}_{\bar{Q}_{n,Y}(W,S)}(W) = E_X\{ \bar{Q}_{n,Y}(W,S) \mid A = a_2, W \}$. 

\begin{theorem}
Assume the following conditions: \begin{itemize}[leftmargin=.3in]
    \item[A1] $\mbox{sup}_{w} | g_{n,A \mid W}(a_1 \mid w) - g_{A \mid W}(a_1 \mid  w) | = o_{\text{p}}(1)$ and $|| g_{n,A \mid W}(a_1 \mid  \cdot) - g_{A \mid W}(a_1 \mid  \cdot) ||_P = o_{\text{p}}(n^{-1/4})$
    \item[A2] $\mbox{sup}_{w} | g_{n,C}(1 \mid a_1, w) - g_{C}(1 \mid a_1, w) | = o_{\text{p}}(1)$ and $|| g_{n,C}(1 \mid a_1,  \cdot) - g_{C}(1 \mid a_1,  \cdot) ||_P = o_{\text{p}}(n^{-1/4})$
    \item[A3] $\mbox{sup}_{w, s} | g_{n,A \mid W, S}(a_1 \mid w, s) - g_{A \mid W, S}(a_1 \mid w, s) | = o_{\text{p}}(1)$ and $|| g_{n,A \mid W, S}(a_1 \mid \cdot) - g_{A \mid W, S}(a_1 \mid \cdot) ||_{P_X} = o_{\text{p}}(n^{-1/4})$
    \item[A4] $|| \bar{Q}_{n,Y} - \bar{Q}_Y ||_{P_X} = o_{\text{p}}(n^{-1/4})$ 
    \item[A5] $|| \bar{Q}_{n,\bar{Q}_Y(W,S)} - \bar{Q}_{\bar{Q}_{n,Y}(W,S)} ||_{P_X} = o_{\text{p}}(n^{-1/4})$ and $|| \bar{Q}_{n,D(P_X)} - \bar{Q}_{D(\hat{P}_X)} ||_P = o_{\text{p}}(n^{-1/4})$
    \item[A6] $|| \hat{D}_1 - D(P) ||_P^2 = o_{\text{p}}(1)$ and $\hat{D}_1$ falls in a $P$-Donsker class with probability tending to 1.
\end{itemize}
Under these assumptions $n^{1/2}\{\psi_{n,1}^+(a_1, a_2) - \psi(a_1, a_2)\}$ converges in distribution to a random variable with a Normal(0, $E\{ D^2(P)(O) \}$) distribution.
\end{theorem}

We provide a proof in the web supplement. The uniform consistency assumptions in A1-A3 ensure estimates of treatment and censoring probabilities are bounded away from zero. The $L^2$ rate conditions in A1-A5 ensure negligibility of a second-order remainder term. The $n^{-1/4}$ convergence rates are sufficient for the results of the theorem, but could also be weakened to conditions involving the product of rates for various combinations of nuisance parameters. We note that the $n^{-1/4}$ rate is slower than the standard parametric rate of $n^{-1/2}$, implying that flexible regression techniques could be used. However, due to the curse of dimensionality, in settings with even moderately large dimensions of $S$ or $W$, the $n^{-1/4}$ rates may be difficult to achieve without further smoothness assumptions. For example, the highly adaptive lasso assumes that the nuisance functions have variation norm bounded by a constant and achieves requisite $L^2$ convergence irrespective of the dimension of the regression \citep{benkeser2016highly}. However, this bounded variation condition can be rather stringent in higher dimensions. Assumption A6 is needed to ensure that an empirical process term is negligible \citep{van1996empirical}, but could be removed by using cross-fitting as in \citet{zheng2011cross} and \citet{chernozhukov2017double}. 

The robustness of the full- and observed-data efficient influence functions imply that $\psi_{n,1}^+(a_1, a_2)$ is consistent if: (i) either (i.a) $\bar{Q}_{n,Y}$ and $\bar{Q}_{n,\bar{Q}_Y(W,S)}$ are consistent for their respective targets {\it or} (i.b) $g_{n, A \mid W, S}$, $g_{n, A \mid W}$, and $g_{n,C}(1 \mid a_1, \cdot)$ are consistent {\it or} (i.c) $\bar{Q}_{n,Y}$, $g_{n, A \mid W}$, and $g_{n,C}(1 \mid a_1, \cdot)$ are consistent {\it and} (ii) either (ii.a) $g_{n,R}$ is consistent for $g_R$ or (ii.b) $\bar{Q}_{n,D(P_X)}$ is consistent for $\bar{Q}_{D(P_X)}$.

Under the assumptions of the theorem, we also have that $\sigma^2_{n,1} = n^{-1} \sum_{i=1}^n \{\hat{D}_1(O_i) - n^{-1} \sum_{j=1}^n \hat{D}_1(O_j) \}^2$ is a consistent estimate of the asymptotic variance of $n^{1/2} \psi_{n,1}^+(a_1,a_2)$ that can be used to construct Wald-type confidence intervals. 

\subsection{An alternative one-step estimator}

Our second estimator differs both in terms of the construction of the plug-in estimator, as well as in the form of the efficient influence function used in the one-step estimator. Both changes are the result of a more explicit examination of $\bar{Q}_{D(P_X)}$ in (\ref{obsdata_eif}). We define \begin{align*}
& \tilde{Q}_{D(P_X)}(W, A, C, CY) := E \left[ \frac{\ind(A = a_1, C = 1)}{g_{A \mid W}(a_2 \mid W) g_{C}(1 \mid a_1, W)} \frac{g_{A \mid W,S}(a_2 \mid W, S)}{g_{A \mid W,S}(a_1 \mid W, S)} \right. \\
&\hspace{3in} \times \{ Y - \bar{Q}_{Y}(W, S) \} \mid R = 1, W, A, C, CY  \bigg] \ ,
\end{align*}
which is similar to $\bar{Q}_{D(P_X)}$, but only involves the first term in $D(P_X)$. We also define $
\tilde{Q}_{\bar{Q}_{Y}(W, S)}(W, A, C, CY) := E\{ \bar{Q}_{Y}(W, S) \mid R = 1, W, A, C, CY \}$, 
and $\tilde{Q}_{\tilde{Q}(W, A, C, CY)}(W) = E\{\tilde{Q}_{\bar{Q}_{Y}(W, S)}(W, A, C, CY) \mid A = a_2, W \}$. In the appendix, we show that $\tilde{Q}_{\tilde{Q}(W, A, C, CY)} = \bar{Q}_{\bar{Q}_Y(W,S)}$ defined above. 

Using the above-defined nuisance parameters, we can rewrite (\ref{obsdata_eif}) as follows, \begin{equation} \label{obsdata_eif_new} \begin{aligned}
    D(P)(o) &:= \frac{r}{g_{R}(w, a, c, cy)} \frac{\ind(a = a_1, c = 1)}{g_{A \mid W}(a_2 \mid w) g_{C}(1 \mid a_1, w)} \frac{g_{A \mid W,S}(a_2 \mid w, s)}{g_{A \mid W,S}(a_1 \mid w, s)} \{ y - \bar{Q}_{Y}(w,s) \} \\
    &\hspace{0.4in} + \frac{r}{g_{R}(w, a, c, cy)} \frac{\ind(a = a_2)}{g_{A \mid W}(a_2 \mid w)} \{\bar{Q}_{Y}(w,s) - \tilde{Q}_{\bar{Q}_{Y}(W, S)}(w, a, c, cy)\} \\
    &\hspace{0.5in} + \frac{\ind(a = a_2)}{g_{A \mid W}(a_2 \mid w)} \{\tilde{Q}_{\bar{Q}_{Y}(W, S)}(w, a, c, cy) - \tilde{Q}_{\tilde{Q}(W, A, C, CY)}(w) \} \\
    &\hspace{0.6in} + \tilde{Q}_{\tilde{Q}(W, A, C, CY)}(w) - \int \tilde{Q}_{\tilde{Q}(W, A, C, CY)}(w) dQ_W(w) \\ 
    &\hspace{0.8in}- \frac{\tilde{Q}_D(w, a, c, cy)}{g_R(1 \mid w, a, c, cy)} \{r - g_R(1 \mid w, a, c, cy)\} \ . 
\end{aligned}
\end{equation}
To construct a one-step estimator based on this representation, we describe construction of a plug-in estimate and an approach for evaluating the one-step correction term.

For the plug-in estimator, we begin as above by obtaining $\bar{Q}_{n,Y}$, an estimate of $\bar{Q}_{Y}$ using inverse weighted regression and evaluating $\bar{Q}_{n,Y}(W_i, S_i)$ for all $i$ such that $R_i = 1$. We then use these estimates as the outcome in a regression onto $W, A, C, CY$ amongst those with $R = 1$, which yields an estimate $\tilde{Q}_{n, \bar{Q}_{Y}(W, S)}$ of $\tilde{Q}_{\bar{Q}_{Y}(W, S)}$. Similarly, we evaluate $\tilde{Q}_{n, \bar{Q}_{Y}(W, S)}(W_i, A_i, C_i, C_i Y_i)$ for $i = 1, \dots, n$ and use this as the outcome in a regression onto $W$ amongst those with $A = a_2$, thereby providing an estimate $\tilde{Q}_{n, \tilde{Q}(W, A, C, CY)}$ of $\tilde{Q}_{\tilde{Q}(W, A, C, CY)}$. Finally, the plug-in estimator is $\psi_{n,2}(a_1, a_2) := n^{-1} \sum_{i=1}^n \tilde{Q}_{n, \tilde{Q}(W, A, C, CY)}(W_i)$. 

With the estimates of the sequential outcome regression described above and the estimates of and $g_R, g_{A \mid W}, g_{A \mid W, S}$ described above, the only additional nuisance parameter estimate we require is one of $\tilde{Q}_D$. Such an estimate can be obtained by evaluating \[
\hat{D}_{X,1}(O_i) := \frac{\ind(A_i = a_1, C_i = 1)}{g_{n, A \mid W}(a_2 \mid W_i) g_{n, C}(1 \mid a_1, W_i)} \frac{g_{n, A \mid W,S}(a_2 \mid W_i, S_i)}{g_{n, A \mid W,S}(a_1 \mid W_i, S_i)} \ \{ Y_i - \bar{Q}_{n, Y}(W_i, S_i) \} 
\]
for all $i$ such that $R_i = 1$. The quantity $\hat{D}_{X,1}(O_i)$ then serves as an outcome in a regression onto $W_i, A_i, C_i, C_i Y_i$, yielding an estimate $\tilde{Q}_{n,D}$ of $\tilde{Q}_D$. 

We denote by $\hat{D}_2$ the efficient influence function in (\ref{obsdata_eif_new}), but with true nuisance parameters substituted with their estimated counterparts, just as we did in (\ref{eval_eif_at_estimates}) for the classic one-step estimator. Our proposed one-step estimator is $\psi_{n,2}^+(a_1, a_2) = \psi_{n,2}(a_1, a_2) + n^{-1} \sum_{i=1}^n \hat{D}_2(O_i)$. We have the following theorem describing its behavior. Let $\tilde{Q}_{D(\hat{P}_X)}(W, A, C, CY) = E\{ \hat{D}_{2,1}(O) \mid W, A, C, CY \}$ and $\tilde{Q}_{\bar{Q}_{n,Y}(W,S)}(W, A, C, CY) := E\{ \bar{Q}_{n,Y}(W,S) \mid R = 1, W, A, C, CY \}$. 

\begin{theorem}
Assume conditions A1-A4 of Theorem 1. Additionally assume \begin{itemize}[leftmargin=0.3in]             
    \item[B5] $||\tilde{Q}_{n,\bar{Q}_{Y}(W, S)} - \tilde{Q}_{\bar{Q}_{n,Y}(W, S)}||_P = o_{\text{p}}(n^{-1/4})$ and $|| \tilde{Q}_{n, \tilde{Q}(W, A, C, CY)} - \tilde{Q}_{\tilde{Q}(W, A, C, CY)} ||_P = o_{\text{p}}(n^{-1/4})$
    \item[B6] $|| \tilde{Q}_{n,D(P_X)} - \tilde{Q}_{D(\hat{P}_X)} ||_P = o_{\text{p}}(n^{-1/4})$
    \item[B7] $|| \hat{D}_2 - D(P) ||_P^2 = o_{\text{p}}(1)$ and $\hat{D}_2$ falls in a $P$-Donsker class with probability tending to 1.
\end{itemize}
Under these assumptions $n^{1/2}\{\psi_{n,2}(a_1, a_2) - \psi(a_1, a_2)\}$ converges in distribution to a random variable with a Normal(0, $E\{ D^2(P)(O) \}$) distribution.
\end{theorem}

The conditions are similar to those of Theorem 1. The estimator $\psi_{n,2}(a_1, a_2)$ is also multiply robust in the sense of the classic one-step estimator, but where we replace (i.a) with condition (i.a') $\bar{Q}_{n,Y}$, $\tilde{Q}_{n, \tilde{Q}(W, A, C, CY)}$ are consistent for their respective targets and replace (ii.b) with condition (ii.b') $\tilde{Q}_{n,D(P_X)}$ is consistent for $\tilde{Q}_{D(P_X)}$ {\it and} $\tilde{Q}_{n,\bar{Q}_Y(W,S)}$ is consistent for $\tilde{Q}_{\bar{Q}_Y(W,S)}$. Under the assumptions of the theorem, $\sigma^2_{n,2} = n^{-1} \sum_{i=1}^n \{\hat{D}_2(O_i) - n^{-1} \sum_{j=1}^n \hat{D}_2(O_j) \}^2$ is a consistent estimate of the asymptotic variance of $n^{1/2} \psi_{n,2}(a_1,a_2)$. 

We provide a detailed theoretical comparison of the two one-step estimators in the supplementary materials.

\section{Practical considerations in the context of COVID vaccines}

We turn to a discussion of the implications of our theoretical results on planning for the analysis of  randomized trials of COVID vaccines. First, since we are in the context of a randomized trial, $g_{A \mid W}$ is known exactly. Thus, the conditions pertaining to consistent estimation of this quantity are easily satisfied. For example, we could use the known randomization probabilities or fit a low-dimensional parametric regression to adjust for chance imbalances in covariates. Similarly, the sampling probability $g_R$ is known by design. If we use the known sampling probability in the estimation procedure, then we can use any estimator of $\bar{Q}_{D(P_X)}$ (or, for the alternative one-step, $\tilde{Q}_{D(P_X)}$) and the results of the theorem hold {\it irrespective} of the consistency of $\bar{Q}_{n,D}$ ($\tilde{Q}_{n,D}$). All that we need is that these estimates converge to {\it something} at a reasonable rate.

Second, we comment on estimation of regression quantities that involve conditioning on $C$ and $CY$. Note that there are three categories into which a participant's data may fall, right-censored ($C = 0, CY = 0$), not right-censored and not a case ($C = 1, CY = 0$), and not right-censored and a case ($C = 1, CY = 1$). Thus, for regression quantities that require conditioning on these data, we can construct dummy variables for two of these cases to include in the regression. 

Third, in the above developments, we have left unspecified which regression techniques may be used to estimate the nuisance parameters. We provide a software implementation that implements two specific approaches, but also that allows users to define their own regression approaches. Specifically, we adopt approaches based on working linear or logistic (depending on the parameter space of the particular nuisance parameter) regression models, as well as based on super learner \citep{van2007super}. In the latter approach, one defines a pre-specified {\it library} of candidate regression estimators. The resultant regression estimator is a convex combination of the library of candidate estimators where the weights are selected by minimizing a cross-validated risk criterion. 

\section{Simulations}

\subsection{Confirming theoretical properties of estimators}

The goal of this simulation was to verify the statistical properties of our estimators as established by our theorems. Accordingly, we used a discrete data generating process so that it was possible to (i) numerically approximate the efficiency bound and (ii) use nonparametric maximum likelihood estimators for nuisance parameters thereby guaranteeing that requisite regularity conditions required by our theorems were satisfied. We generated data as follows. A bivariate variable $W = (W_1, W_2)$ was generated by independently drawing two Bernoulli(1/2) variables. Given $W = w$, a binary treatment $A$ was generated according to a Bernoulli distribution with $g_{A \mid W}(1 \mid w) = \mbox{expit}(w_1 - w_2)$. Given $W = w$ and $A = a$, $S$ was drawn from a Binomial(2, $p(a,w)$) distribution with $p(a,w) = \mbox{expit}(-1 + w_1 / 4  - w_2 / 3 + a / 2)$. Given $W = w, A = a, S = s$, a binary outcome $Y$ was drawn from a Bernoulli distribution with $P(Y = 1 \mid W = w, A = a, S = s) = \mbox{expit}(-2 + a / 2 + w_1 / 2 - s / 2)$, while $C$ was drawn from a Bernoulli distribution with $g_C(1 \mid a, w) = \mbox{expit}(2 + w_1 / 2 - w_2 / 3)$. Subsequently, case-cohort sampling was applied to $S$, by sampling a random 1/4 of the cohort, in addition to all observations with $CY = 1$. We evaluated our two proposed estimators of $\psi(1,0)$, the true value of which was approximately 0.187 with efficient variance bound equal to $E\{D^2(P)(O) \}$ = 0.509.

For each sample size $n \in \{500, 1000, 2000, 4000, 8000\}$, we simulated 1000 data sets from this data generating process and computed our two proposed estimators along with a confidence interval for $\psi(1,0)$. We evaluated the estimators in terms of their bias (scaled by $n^{1/2}$), their standard error (scaled by $n^{1/2}$), the coverage probability of a nominal 95\% confidence interval, and the ratio of the scaled standard error to the square root of the efficient variance. 

We first evaluated estimators under the conditions of the theorem where all nuisance parameters are consistently estimated at appropriate rates. To achieve this, we used parsimonious logistic regression models for $g_{A \mid W}$, $g_{C}$, $g_R$, $\bar{Q}_Y$ and fully saturated logistic regression models (i.e., nonparametric maximum likelihood estimates) for the remaining nuisance parameters. In this setting, our theory predicts that bias of the estimators should converge to zero faster than $n^{-1/2}$, the scaled standard error of the estimators should converge to the square root of the efficiency bound, and confidence intervals should attain nominal coverage. 

After confirming asymptotic properties in the setting with consistently estimated nuisance parameters, we subsequently studied six scenarios in which various combinations of nuisance parameters were misspecified. The six scenarios that we studied are ones in which our multiple robustness results imply that the one-step estimators should remain consistent. In these settings, certain nuisance parameters were inconsistently estimated by using a simple intercept-only regression model (i.e., the sample average of the outcome of the regression). In these settings, our results suggest that we should still see diminishing bias and stabilizing scaled standard error as sample size increases. However, neither  estimators' influence function will equal the efficient influence function. Consequently, we expect non-nominal coverage probabilities for the confidence intervals.

The results of the simulation results confirmed our theory and estimators enjoyed expected behavior in all settings considered (Table \ref{sim1_rslt}). In the setting where all nuisance parameters were consistently estimated, the two one-step estimators had similar performance across the board. However, in settings where the regression of the full data efficient influence function was inconsistently estimated, the alternative one-step estimator $\psi_{n,2}^+(1,0)$ tended to have higher variability.

\begin{table}[ht]
\centering
\caption{Bias (scaled by $n^{1/2}$), standard error ((scaled by $n^{1/2}$, SE), coverage probability of a nominal 95\% confidence interval (Cov), and the scaled SE divided by the square root of the efficient variance (Ratio) for the two proposed one-step estimators in different settings. The settings column indicates which nuisance parameters are consistently estimated ($g_R$ was consistently estimated in all settings).}
\label{sim1_rslt}
\begin{tabular}{|r|rrrrr|rrrr}
   & & \multicolumn{4}{|c|}{$\psi_{n,1}^+(1,0)$} & \multicolumn{4}{|c|}{$\psi_{n,2}^+(1,0)$}\\
 \hline Setting & $n$ & Bias & SE & Cov. & Ratio & Bias & SE & Cov. & Ratio \\ \hline
 \multirow{5}{6em}{All} & 500 & 0.33 & 1.06 & 0.93 & 1.48 & 0.37 & 1.11 & 0.94 & 1.55 \\ 
 & 1000 & 0.21 & 0.83 & 0.95 & 1.17 & 0.21 & 0.83 & 0.95 & 1.16 \\ 
   & 2000 & 0.14 & 0.79 & 0.94 & 1.11 & 0.14 & 0.79 & 0.94 & 1.11 \\ 
   & 4000 & 0.10 & 0.74 & 0.94 & 1.04 & 0.10 & 0.74 & 0.94 & 1.04 \\ 
   & 8000 & 0.04 & 0.73 & 0.95 & 1.02 & 0.04 & 0.73 & 0.95 & 1.02 \\ 
   \hline
\multirow{5}{6em}{$\bar{Q}_Y$, $\bar{Q}_{\bar{Q}_Y}$} & 500 & 0.03 & 0.74 & 0.99 & 1.03 & 0.14 & 1.05 & 0.98 & 1.47 \\ 
   & 1000 & 0.03 & 0.74 & 0.99 & 1.04 & 0.10 & 1.06 & 0.98 & 1.49 \\ 
   & 2000 & -0.00 & 0.72 & 0.99 & 1.01 & 0.03 & 1.04 & 0.98 & 1.46 \\ 
   & 4000 & 0.05 & 0.75 & 0.99 & 1.04 & 0.10 & 1.12 & 0.97 & 1.57 \\ 
   & 8000 & 0.01 & 0.77 & 0.99 & 1.08 & 0.02 & 1.11 & 0.97 & 1.56 \\ 
   \hline
\multirow{5}{6em}{$\bar{Q}_Y$, $\bar{Q}_{\bar{Q}_Y}$, $\bar{Q}_{D_{P_X}}$, $\tilde{Q}_{D_{P_X}}$} & 500 & -0.01 & 0.67 & 0.90 & 0.94 & -0.01 & 0.67 & 0.90 & 0.94 \\ 
   & 1000 & -0.01 & 0.68 & 0.92 & 0.95 & -0.01 & 0.68 & 0.92 & 0.95 \\ 
   & 2000 & 0.01 & 0.66 & 0.92 & 0.93 & 0.01 & 0.67 & 0.92 & 0.93 \\ 
   & 4000 & 0.04 & 0.63 & 0.93 & 0.89 & 0.04 & 0.63 & 0.93 & 0.89 \\ 
   & 8000 & -0.01 & 0.65 & 0.93 & 0.91 & -0.01 & 0.65 & 0.93 & 0.91 \\ 
   \hline
\multirow{5}{6em}{$g_{A \mid W}$, $g_{A \mid W, S}$, $g_C$} & 500 & 0.56 & 1.14 & 0.99 & 1.60 & 0.66 & 1.63 & 0.96 & 2.29 \\ 
   & 1000 & 0.37 & 0.98 & 0.99 & 1.38 & 0.45 & 1.44 & 0.96 & 2.02 \\ 
   & 2000 & 0.21 & 0.88 & 0.99 & 1.24 & 0.25 & 1.29 & 0.97 & 1.81 \\ 
   & 4000 & 0.11 & 0.87 & 0.99 & 1.23 & 0.14 & 1.27 & 0.96 & 1.79 \\ 
   & 8000 & 0.11 & 0.85 & 0.99 & 1.19 & 0.12 & 1.20 & 0.97 & 1.69 \\ 
   \hline
\multirow{5}{6em}{$g_{A \mid W}$, $g_{A \mid W, S}$, $g_C$, $\bar{Q}_{D_{P_X}}$, $\tilde{Q}_{D_{P_X}}$} & 500 & 0.42 & 1.17 & 0.92 & 1.64 & 0.44 & 1.11 & 0.92 & 1.55 \\ 
   & 1000 & 0.30 & 0.87 & 0.94 & 1.22 & 0.31 & 0.87 & 0.93 & 1.22 \\ 
   & 2000 & 0.20 & 0.78 & 0.93 & 1.10 & 0.20 & 0.78 & 0.93 & 1.09 \\ 
   & 4000 & 0.13 & 0.72 & 0.94 & 1.01 & 0.13 & 0.72 & 0.94 & 1.01 \\ 
   & 8000 & 0.13 & 0.76 & 0.92 & 1.06 & 0.13 & 0.76 & 0.92 & 1.06 \\ 
   \hline
\multirow{5}{6em}{$\bar{Q}_Y$, $g_{A \mid W}$, $g_C$} & 500 & 0.04 & 0.84 & 1.00 & 1.17 & 0.13 & 1.31 & 0.98 & 1.84 \\ 
   & 1000 & 0.06 & 0.82 & 1.00 & 1.15 & 0.14 & 1.23 & 0.99 & 1.73 \\ 
   & 2000 & 0.06 & 0.81 & 1.00 & 1.13 & 0.12 & 1.21 & 0.99 & 1.69 \\ 
   & 4000 & 0.03 & 0.80 & 1.00 & 1.12 & 0.08 & 1.20 & 0.99 & 1.69 \\ 
   & 8000 & 0.04 & 0.80 & 1.00 & 1.12 & 0.09 & 1.21 & 0.99 & 1.70 \\ 
   \hline
\multirow{5}{6em}{$\bar{Q}_Y$, $g_{A \mid W}$, $g_C$, $\bar{Q}_{D_{P_X}}$, $\tilde{Q}_{D_{P_X}}$} & 500 & -0.05 & 0.79 & 0.96 & 1.10 & -0.06 & 0.78 & 0.96 & 1.09 \\ 
   & 1000 & -0.05 & 0.73 & 0.97 & 1.02 & -0.06 & 0.72 & 0.97 & 1.01 \\ 
   & 2000 & -0.01 & 0.71 & 0.97 & 0.99 & -0.01 & 0.71 & 0.97 & 0.99 \\ 
   & 4000 & -0.03 & 0.73 & 0.97 & 1.02 & -0.03 & 0.73 & 0.97 & 1.02 \\ 
   & 8000 & 0.00 & 0.72 & 0.97 & 1.01 & 0.00 & 0.72 & 0.97 & 1.01 \\ 
   \hline
\end{tabular}
\end{table}

\subsection{Evaluating estimators in the context of COVID vaccine trials}

In our second simulation, we wished to study the finite-sample performance of our estimators in a setting similar to what might be expected in a correlates study of a COVID-19 vaccine. To that end, we developed a data generating process meant to mimic the expected data outputs of a large ($n = 30000$), placebo-controlled trial. We simulated a three-dimensional covariate $W = (W_1, W_2, W_3)$ with three Bernoulli-distributed components with success probabilities of 2/5, 1/4, 1/4, meant to represent measurements of participants' age ($\ge$ 65 vs. $<$ 65 years old), racial/ethnic minority status (yes vs. no), and presence of COVID-19 risk factors (yes vs. no). For each simulated participant, we drew their randomization assignment $A$ from a Bernoulli(1/2) distribution. A continuous immune marker was then generated as follows. Given $W_1 = w_1$, we drew $S^*$ from a Normal(2 - 0.5 $w_1$, 1) distribution. Next, given $A = a$, we set $S = a \ind(S^* > 0) S^*$. That is, placebo recipients uniformly have $S = 0$, which is reasonable as the primary analysis of COVID-19 trials focuses on previously uninfected individuals. For individuals in the vaccine arm, there are some ``vaccine failures,'' that is individuals in whom the vaccine fails to induce a positive immune response. Under our simulation setting, the probability of failure increases with age. Given $W = w, S = s, A = a$ we set the probability of having an observed COVID-19 disease endpoint as $\mbox{expit}(\alpha - 0.5 s - 1.8 a + 0.2 w_1 + 0.1 w_2 + 0.7 w_3)$. Note that the parameter $\alpha$ controls the total number of breakthrough endpoints. We considered performance of our estimators under several values of $\alpha$ that may be realistic for COVID vaccine efficacy studies. Under this data generating process, the vaccine has high efficacy, as was observed with the first reported COVID-19 vaccines. For simplicity, we assumed there was no censoring of endpoints. The corresponding true parameter values for each setting we considered is shown in Table \ref{sim2_true_rslt}. 

Two-phase sampling of $S$ was simulated according to the sampling plan developed for all Phase 3 trials run in the United States \citep{ows_sampling_plan}. In this design, a stratified random sample is drawn based on the sixteen strata defined by $W$ and $A$. Specifically, 113 vaccine recipients and 15 placebo recipients are sampled from each of the eight strata defined by $W$. All observed COVID-19 endpoints are additionally sampled. 

We were interested primarily in comparing the two one-step estimators in terms of their performance at realistic sample sizes in terms of point estimation and confidence interval coverage for the indirect effect $\psi(1,1) / \psi(1,0)$ and for the proportion mediated, defined as $1 - \mbox{log}\{\psi(1,0) / \psi(0,0) \} / \mbox{log}\{\psi(1,1) / \psi(0,0) \}$, as these are two key parameters for establishing vaccine correlates. For each setting described above, we simulated 1000 data sets and computed the two one-step point estimates and confidence intervals. 

We repeated the simulation under several different modeling strategies for nuisance parameters. The first considered regression models that included all possible interactions between regressors, where $S$ was modeled as a linear term. In this approach, we do not expect that the nuisance parameters are consistently estimated, though the bias from model misspecification is expected to be relatively small. Our second strategy for nuisance parameter estimation considered using the super learner. In our analysis, we used four different regressions that included a main-terms generalized linear model, a stepwise generalized linear model that included all two-way interaction terms, an intercept-only model, and polynomial multivariate adaptive regression splines. Due to the added flexibility of the ensemble model, we believe this approach should come closer to consistent estimation of the nuisance parameters. However, it is a highly data-driven and so we wished to evaluate whether it could be appropriately employed at these sample sizes. Finally, we considered using simple main-terms regression models that did not include any interactions between variables. These models were expected to have the worst performance in modeling the nuisance parameters and served as a ``worst-case scenario'' in our simulation. We report on the generalized linear models with interaction and super learner results in the body and describe the misspecified main terms models in the supplement.

\begin{figure}[H]
\centering
\includegraphics[width=\textwidth]{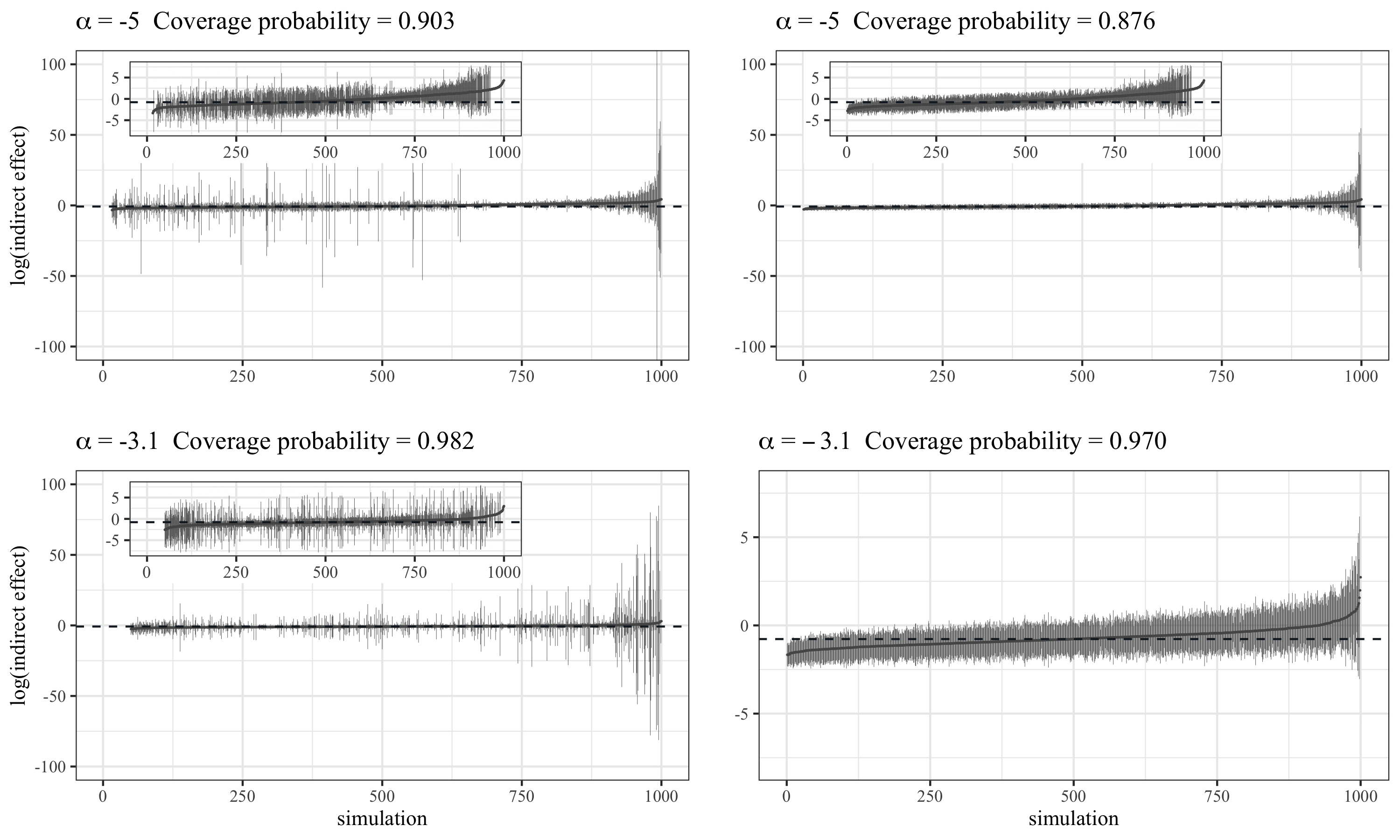}
\caption{Indirect effect confidence intervals using classic (left column) and alternative (right column) one-step estimators constructed using super learner to estimate nuisance quantities. The confidence intervals are ordered across the 1000 simulations from smallest to largest and are displayed on the log scale. In the rare event setting ($\alpha = -5$), we include a subfigure that is zoomed in and has the extreme confidence intervals removed.}
\label{fig:indirect_effect_sl}
\end{figure}

\begin{figure}[H]
\centering
\includegraphics[width=\textwidth]{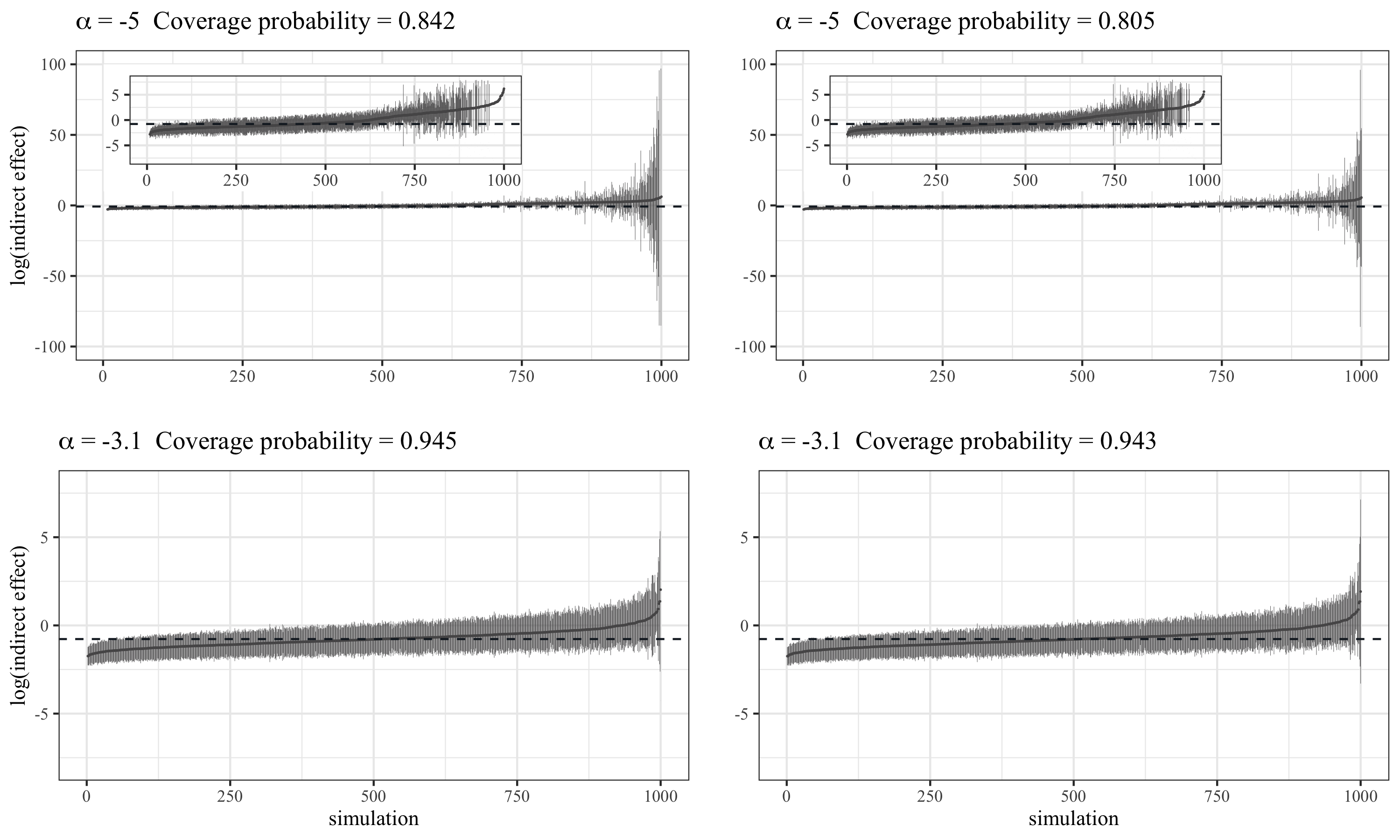}
\caption{Indirect effect confidence intervals using classic (left column) and alternative (right column) one-step estimators constructed using GLMs with interactions to estimate nuisance quantities. The confidence intervals are ordered across the 1000 simulations from smallest to largest and are displayed on the log scale. In the rare event setting ($\alpha = -5$), we include a subfigure that is zoomed in and has the extreme confidence intervals removed.}
\label{fig:indirect_effect_glm_inter}
\end{figure}

For both one-step estimators and irrespective of the nuisance parameter estimation approach, we found that when events were very rare in the vaccine group, the estimators exhibited considerable bias. In these rare-event settings, confidence intervals for the indirect effect and proportion mediated had slightly less than nominal coverage. However, as the number of expected events amongst the vaccinated rose to around 50, the bias of the estimators decreased to a reasonable level and the confidence intervals attained nominal coverage probabilities. Comparing the two modeling strategies, we found comparable results between the GLM with interactions and the SuperLearner. Comparing the two one-step estimators, we saw similar performance in terms of bias and confidence interval coverage. However, we found that confidence intervals for the standard one-step estimator when combined with super learner tended to be more erratic (Figure \ref{fig:indirect_effect_sl}). The same phenomenon was not observed when the GLM with interactions was used to estimate nuisance parameters (Figure \ref{fig:indirect_effect_glm_inter}).

\begin{table}[!htbp]
\centering
\caption{True values of the parameters of interest for the various simulation settings and the average (standard deviation) number of observed events amongst the simulated vaccine recipients ($\bar{n}_1$) and amongst the simulated placebo recipients. VE = 1 - $\psi(1,1) / \psi(0,0)$.}
\label{sim2_true_rslt}
\begin{tabular}{rrrrrll}
  \hline
$\alpha$ & VE & $\frac{\psi(1,1)}{\psi(1,0)}$ & $\frac{\psi(1,0)}{\psi(0,0)}$ & Prop. Mediated & $\bar{n}_1(sd)$ & $\bar{n}_2(sd)$ \\ 
  \hline
-5.00 & 0.92 & 0.46 & 0.17 & 0.30 & 10.6(3.2) & 140.1(11.8) \\ 
  -4.10 & 0.92 & 0.46 & 0.17 & 0.30 & 26.2(5.3) & 339.2(18.0) \\ 
  -3.60 & 0.92 & 0.46 & 0.17 & 0.31 & 43.2(6.6) & 550.3(23.2) \\ 
  -3.30 & 0.92 & 0.46 & 0.17 & 0.31 & 58.4(7.5) & 731.3(26.2) \\ 
  -3.10 & 0.92 & 0.46 & 0.17 & 0.31 & 71.1(8.2) & 882.9(28.9) \\ 
   \hline
\end{tabular}
\end{table}

\begin{table}[!htbp]
\centering
\caption{Bias and confidence interval coverage for the classic (left) and alternative (right) one-step estimator of indirect effect and proportion mediated for different event rates ($\alpha$). Results are shown for when nuisance parameters are estimated using a GLM with interactions (top) and a Super Learner (bottom)}
\resizebox{\columnwidth}{!}{%
\begin{tabular}{rrrrrrrrrl}
  & \multicolumn{4}{c}{$\psi_{n,1}^{+}$} & \multicolumn{4}{c}{$\psi_{n,2}^{+}$} & \\ 
 \hline
  & \multicolumn{2}{c}{Indirect} & \multicolumn{2}{c}{Prop. Med.} & \multicolumn{2}{c}{Indirect} & \multicolumn{2}{c}{Prop. Med.} & \\ 
 $\alpha$ & Bias & Coverage & Bias & Coverage & Bias & Coverage & Bias & Coverage & Method \\ 
 \hline
-5.0 & 4.361 & 0.842 & -0.270 & 0.857 & 3.509 & 0.805 & -0.256 & 0.823 & \multirow{5}{6em}{GLM inter.} \\ 
  -4.1 & 1.103 & 0.922 & -0.098 & 0.926 & 1.807 & 0.911 & -0.100 & 0.916 &  \\ 
  -3.6 & 0.828 & 0.935 & -0.048 & 0.938 & 0.390 & 0.930 & -0.042 & 0.933 & \\
  -3.3 & 0.075 & 0.947 & -0.023 & 0.947 & 0.850 & 0.943 & -0.024 & 0.948 & \\
  -3.1 & 0.025 & 0.945 & -0.013 & 0.950 & 0.083 & 0.943 & -0.013 & 0.948 & \\
   \hline
-5.0 & -0.843 & 0.903 & -0.202 & 0.911 & 1.528 & 0.876 & -0.192 & 0.889 & \multirow{5}{6em}{SuperLearner} \\ 
  -4.1 & 0.396 & 0.947 & -0.079 & 0.947 & 0.375 & 0.927 & -0.065 & 0.931 & \\ 
  -3.6 & 0.261 & 0.977 & -0.052 & 0.979 & 0.187 & 0.959 & -0.035 & 0.960 & \\
  -3.3 & -0.677 & 0.981 & -0.039 & 0.982 & 0.139 & 0.963 & -0.030 & 0.967 & \\
  -3.1 & 0.069 & 0.982 & -0.030 & 0.983 & 0.122 & 0.970 & -0.027 & 0.971 & \\
  \hline
\end{tabular}%
}
\end{table}

We found that basing inference on the misspecified main terms models tended to lead to estimators with  non-negligible bias and conservative inference, with confidence interval coverage probabilities near to 1 (Supplemental material). However, we did find more robust performance of the alternative one-step estimator relative to its standard counterpart. 

Overall, we found that the estimators performed reasonably well in all but the smallest sample sizes considered. We also found that using super learning to estimate nuisance parameters appears acceptable even in settings with a limited number of observed events.

The supplemental material contains additional simulation results for operating characteristics of the estimators of the counterfactual risk parameters $\psi(a_1, a_2)$. Of note, we show there that the classic one-step estimator $\psi_{n,1}^+(1,0) < 0$ for a non-negligible fraction of simulations. The results are particularly bad for the misspecified nuisance models. However, even with super learner $\psi_{n,1}^+(1,0)$ was less than zero up to 5\% of simulated data sets. On the other hand, the alternative one-step estimator was always positive. 
\section{Discussion}

A central contribution of this study is establishing that nonparametric estimators of mediation effects can reasonably be employed in the analysis of COVID-19 correlates data assuming a reasonable number ($\approx$ 50) of vaccine breakthrough cases are observed. The fact that super learner can be used to estimate nuisance parameters while retaining reasonable operating characteristics is highly appealing given (i) the inherent complexity of some of the nuisance parameters that need to be estimated as part of our procedure (e.g., $\tilde{Q}_{D(P_X)}$) and (ii) our limited understanding of the interaction between patient-level COVID-19 risk factors and vaccine-induced immune responses on risk of COVID-19. 

Beyond this practical contribution, we believe that the alternative approach outlined to one-step estimation is also a valuable contribution to the literature on estimation of pathwise differentiable parameters in two-phase sampling designs. An attractive feature of this alternative approach is the decoupling of estimation of the sampling probabilities and estimation of full-data nuisance parameters. This should provide more practical avenues for achieving a robust estimator in settings where sampling probabilities are unknown. However, the results of the simulations were mixed in terms of the performance of this alternative one-step estimator vs. the standard one. In our COVID-19-oriented simulation, we saw more stable inference in small samples using the alternative one-step. The alternative estimator of $\psi(1,0)$ never slipped below 0, while this occurred with some frequency for the classic estimator. However, in the large sample simulation, we saw that the alternative one-step may not be as robust to certain forms of nuisance parameter misspecification. A careful comparison of these estimation approaches across a more diverse set of scenarios is warranted and is an important direction for future research. 

Another potential area for expanding this work area is in developing a targeted minimum loss-based estimation (TMLE) framework for estimation of these mediation parameters. As substitution estimators, TMLEs enjoy the property of being guaranteed to fall in the parameter space. This property may be particularly important in the context of studies of highly effective COVID-19 vaccines where risks of disease amongst the vaccinated are extremely small. Very modest one-step correction terms can easily push estimators below zero, as we saw in the simulation study. TMLEs would likely enjoy more robust behavior in these settings.

\section*{Data Availability Statement}
The code needed to reproduce the both simulation studies is available on in a GitHub repository at https://github.com/benkeser/natmed2\_sims. 

\bibliographystyle{apalike}
\bibliography{refs}

\section*{Supporting Information}
The \texttt{R} package \texttt{natmed2} is available on Github at https://github.com/benkeser/natmed2. Web Appendices referenced appear below. Appendix~A includes a proof of Theorems 1 and 2;  Appendix B contains theoretical comparison of the classic and alternative one-step approaches; Appendix C provides an alternative identification result that does not rely at all on full-data nuisance parameters; and Appendix D contains additional simulation results.

\section*{A. Proof of asymptotic linearity}
For proving asymptotic linearity of the proposed estimators, it will first be useful to establish the following lemmas. We recall the following definitions: $\bar{Q}_{\bar{Q}_Y(W,S)}(W) = E_X\{ E_X(Y \mid A = a_1, W, S) \mid A = a_2, W \}$, $\tilde{Q}_{\bar{Q}_Y(W,S)}(W,A,C,CY) = E\{ E_X(Y \mid A = a_1, W, S) \mid R = 1, W, A, C, CY \}$, and $\tilde{Q}_{\tilde{Q}(W,A,C,CY)}(W) = E\{\tilde{Q}_{\bar{Q}_Y(W,S)}(W,A,C,CY) \mid A = a_2, W\}$. 

\begin{lemma}\label{lemma:equiv_params}
We have $\bar{Q}_{\bar{Q}_Y(W,S)} = \tilde{Q}_{\tilde{Q}(W,A,C,CY)}$. 
\end{lemma}

\begin{proof}
\begin{align*}
    \bar{Q}_{\bar{Q}_Y(W,S)}(W) &= E_X\{\bar{Q}_Y(W,S) \mid A = a_2, W \} \\ 
    &= E_X[E_X\{\bar{Q}_Y(W,S) \mid W, A, C, CY \} \mid A = a_2, W] \\
    &= E_X[E\{\bar{Q}_Y(W,S) \mid R = 1, W, A, C, CY \} \mid A = a_2, W] \\
    &= E_X[\tilde{Q}_{\bar{Q}_{Y}(W, S)}(W, A, C, CY) \mid A = a_2, W] \\
    &= E[\tilde{Q}_{\bar{Q}_{Y}(W, S)}(W, A, C, CY) \mid A = a_2, W] \\
    &= \tilde{Q}_{\tilde{Q}(W, A, C, CY)}(W)
\end{align*}
The first line follows by definition of $\bar{Q}_{Y}$, the second by the tower rule. The third follows from the fact that two-phase sampling status $R$ is independent of $\bar{Q}_Y(W,S)$ given $(W, A, C, CY)$. The fourth follows again by definition, the fifth recognizing that the conditional distribution of $(A, C, CY)$ given $A = a_2, W$ implied by $P_X$ is the same as that implied by $P$. The final line follows by definition. 
\end{proof}

\begin{lemma}\label{lemma:rem_full}
Let $\mathcal M^F$ denote the non-parametric  full-data model described in the main document. Let $\Psi^F:\mathcal M^F\rightarrow \mathbb R$ denote the parameter mapping that maps any distribution $P_X'$ in the model to a number $E_{X}'[ E_{X}' \{E_{X}'(CY \mid A = a_1, C = 1, S, W) \mid A = a_2, W\}]$, where $E_X'$ denotes expectation under $P_X'$. The efficient influence function $D(P_X)$ of $\Psi$ in the full data model satisfies
\[\Psi^F(P_X') - \Psi^F(P_X) = - E_{X} \{D(P_X')(X)\} + R_2^F(P_X', P_X),\]
where $R_2^F = R_{2,1}^F + R_{2,2}^F + R_{2,3}^F$ with
\begin{align*}
    &R_{2,1}^F(P_X', P_X) := E_X\left[\{\bar{Q}'_{\bar{Q}'_Y(W,S)}(W)- \bar{Q}_{\bar{Q}_Y(W,S)}(W)\}\left\{1 - \frac{g_{A \mid W}(a_2 \mid W)}{g'_{A \mid W}(a_2 \mid W)}\right\}\right]\\
     &R_{2,2}^F(P_X', P_X) \\
     &\hspace{0.05in} := E_X \bigg[ \frac{g_{A \mid W}(a_2 \mid W)}{g'_{A \mid W}(a_2 \mid W)}\left\{\frac{g'_{A \mid W,S}(a_2 \mid W, S)}{g'_{A \mid W,S}(a_1 \mid W, S)}-\frac{g_{A \mid W,S}(a_2 \mid W, S)}{g_{A \mid W,S}(a_1 \mid W, S )}\right\} \{\bar{Q}_{Y}(W, S) - \bar{Q}'_{Y}(W, S)\} \bigg] \\
         &R_{2,3}^F(P_X', P_X) := E_X\left[\frac{g'_{A \mid W,S}(a_2 \mid W, S)}{g'_{A \mid W,S}(a_1 \mid W, S)}\left\{\frac{g_C(1\mid a_1, W)}{g'_C(1\mid a_1, W)}-1\right\}\{\bar{Q}_{Y}(W, S) - \bar{Q}'_{Y}(W, S)\}\right]\ .
\end{align*}
\end{lemma}
\begin{proof}
This lemma follows from Theorem 4 in the supplementary materials of \cite{diaz2019non}.
\end{proof}
\begin{lemma} \label{lemma:rem_obs}
Let $V=(W, A, C, CY)$, and let $P'$ denote the observed data distribution implied by coarsening $P_X'$ according to $g_R'$. The efficient influence function $D(P)$ of $\Psi$ in the observed data model satisfies
\[\Psi^F(P_X') - \Psi^F(P_X) = - E \{D(P')(O)\} + R_2(P_X', P_X, P', P),\]
where
\begin{align}
   &R_2(P_X', P_X, P', P)= R_2^F(P_X', P_X) \notag \\
   &\hspace{0.1in} +E\left(\left\{\frac{g_R(1 \mid V)-g_R'(1 \mid V)}{g_R'(1 \mid V)}\right\}\left[E\{D(P_X')(O)\mid R=1, V\} - E'\{D(P_X')(O)\mid R=1, V\}\right] \right) \label{obs_data_remainder}
\end{align}

\end{lemma}
\begin{proof}
This follows from Lemma~\ref{lemma:rem_full} and the definition of $D(P)$ after computing 
\[ R_2(P', P)=\Psi^F(P_X') - \Psi^F(P_X)+ E \{D(P')(O)\},\]
using the assumption $R\independent X\mid V$.
\end{proof}

\begin{corollary}
The object $R_{2,1}^F$ can equivalently be written as \begin{align*}
R_{2,1}^F(P_X', P_X) &= E_X\left[\{\tilde{Q}'_{\tilde{Q}'(W,A,C,CY)}(W) - \tilde{Q}_{\tilde{Q}(W,A,C,CY)}(W)\}\left\{1 - \frac{g_{A \mid W}(a_2 \mid W)}{g'_{A \mid W}(a_2 \mid W)}\right\}\right] \ , 
\end{align*}
and the object $R_2 = R_2^F+ R_{2,1} + R_{2,2}$ where \begin{align*}
    &R_{2,1}(P_X', P_X, P', P) = E \left[ \left\{\frac{g_R(1 \mid V) - g_R'(1 \mid V)}{g_R'(1 \mid V)} \right\} \{\tilde{Q}_{D(P_X')}(V) - \tilde{Q}'_{D(P_X')}(V) \} \right] \\
    &R_{2,2}(P_X', P_X, P', P) \\
    &\hspace{0.1in} = E \left[ \frac{R}{g_R'(1 \mid V)} \frac{\ind(A = a_2)}{g'_{A \mid W}(a_2 \mid W)} \left\{ \frac{g_R(1 \mid V) - g_R'(1 \mid V)}{g_R'(1 \mid V)} \right\} \{\tilde{Q}_{\bar{Q}_Y'(W,S)}(V) - \tilde{Q}'_{\bar{Q}_Y'(W,S)}(V) \} \right] \ . 
\end{align*}
\end{corollary}

\begin{proof}
The equality of $R_{2,1}^F$ follows immediately from Lemma \ref{lemma:equiv_params}. The equality of $R_2$ comes from an algebraic simplification of (\ref{obs_data_remainder}) in Lemma \ref{lemma:rem_obs}.
\end{proof}

We are now ready to prove the theorems. We start with the proof of asymptotic linearity for the classic one-step estimator. Below, we use $\hat{P}$ and $\hat{P}_X$ denote any distribution in the observed and full data models, respectively, that is compatible with our estimated nuisance parameters. We define the shorthand notation $g_j(W) := g_{A\mid W}(a_j \mid W)$ and $g_{n,j}(W) := g_{n, A \mid W}(a_j \mid W)$. Similarly, we define $\tilde{g}_{j}(W,S) := g_{A\mid W, S}(a_j \mid W, S)$, $\tilde{g}_{n, j}(W,S) := g_{n, A\mid W, S}(a_j \mid W, S)$, and $g_{C,j}(W) := g_C(1\mid a_j, W)$, $g_{n,C,j}(W) := g_{n,C}(1\mid a_j, W)$. We also make use of the shorthand notation $P' f$ to denote expectation of a $P'$-measurable function $f$ of $O$ under a given probability distribution $P'$; thus, $P'f = \int f(o) dP(o)$. In particular, we denote by $P_n$ the empirical distribution based on $O_1, \dots, O_n$ so that $P_n f = n^{-1} \sum_{i=1}^n f(O_i)$. Note that based on these definitions, the classic one-step estimator can be written as $\psi_{n,1}^+(a_1, a_2) = \Psi^F(\hat{P}_X) + P_n D(\hat{P}_X)$. Thus, by Lemma \ref{lemma:rem_obs}, \begin{align*}
    \psi_{n,1}^+(a_1, a_2) - \psi(a_1, a_2) &= (P_n - P) D(\hat{P}) + R_2(\hat{P}_X, P_X, \hat{P}, P) \\
    &= (P_n - P) D(P) + R_2(\hat{P}_X, P_X, \hat{P}, P) + o_{\text{p}}(n^{-1/2}) \\
    &= P_n D(P) + R_2(\hat{P}_X, P_X, \hat{P}, P) + o_{\text{p}}(n^{-1/2})
\end{align*}
which follows from assumption A6 and the fact that $D(P)$ has mean zero. Thus, it remains to establish that $R_2(\hat{P}_X, P_X, \hat{P}, P) = o_{\text{p}}(n^{-1/2})$. We note that \begin{align*}
    R_{2,1}^F(\hat{P}_X, P_X) &= P \left[ g_{n,A\mid W}^{-1}(a_2 \mid \cdot) \{\bar{Q}_{n, \bar{Q}_{n,Y}(W,S)} - \bar{Q}_{\bar{Q}_Y(W,S)}\} \{ g_{n,A\mid W}(a_2 \mid \cdot) - g_{A\mid W}(a_2 \mid \cdot) \} \right] \\
    &\le \left\{ \underset{w}{\mbox{sup}} \ g_{n,A\mid w}^{-1} \right\} | P \left[ \{\bar{Q}_{n, \bar{Q}_{n,Y}(W,S)} - \bar{Q}_{\bar{Q}_Y(W,S)}\} \{ g_{n,A\mid W}(a_2 \mid \cdot) - g_{A\mid W}(a_2 \mid \cdot) \} \right] | \\
    &\le \left\{ \underset{w}{\mbox{sup}} \ g_{n,A\mid W}^{-1} \right\} || \bar{Q}_{n, \bar{Q}_{n,Y}(W,S)} - \bar{Q}_{\bar{Q}_Y(W,S)} ||_P || g_{n,A\mid W}(a_2 \mid \cdot) - g_{A\mid W}(a_2 \mid \cdot) ||_P \\
    &= o_{\text{p}}(n^{-1/2}) \ ,
\end{align*}
where the last line follows from assumptions A1 and A5. Next, we note that $R_{2,2}^F(P_X', P_X)$ can be written as \begin{align*}
    P_X \left[\frac{g_2}{g_{n,2}} \left\{ \frac{(\tilde{g}_{1} - \tilde{g}_{n,1}) + (\tilde{g}_{n,2} - \tilde{g}_{2}) + \tilde{g}_2(\tilde{g}_{1} - \tilde{g}_{n,1}) + \tilde{g}_{n,1}(\tilde{g}_{n,2} - \tilde{g}_{2})}{\tilde{g}_{n,1}\tilde{g}_{1}} \right\} (\bar{Q}_{Y} - \bar{Q}_{n,Y}(W, S)) \right] \ . 
\end{align*}
Thus, $R_{2,2}^F(\hat{P}_X, P_X)$ can be split into four terms that each may be analyzed separately in a similar fashion as with $R_{2,1}^F$. For example, we have \begin{align*}
    &P_X \left[\frac{g_2}{g_{n,2}} \left\{ \frac{\tilde{g}_{1} - \tilde{g}_{n,1}}{\tilde{g}_{n,1}\tilde{g}_{1}} \right\} (\bar{Q}_{Y} - \bar{Q}_{n,Y}) \right] \\ 
    &\hspace{0.2in} \le \left(\underset{w,s}{\mbox{sup}} \ \frac{g_2(w)}{g_{n,2}(w)\tilde{g}_{n,1}(w,s)\tilde{g}_{1}(w,s)} \right) | P_X\left\{ (\tilde{g}_{1} - \tilde{g}_{n,1}) ( \bar{Q}_{Y} - \bar{Q}_{n,Y}) \right\} | \\
    &\hspace{0.2in} \le \left(\underset{w,s}{\mbox{sup}} \ \frac{g_2(w)}{g_{n,2}(w)\tilde{g}_{n,1}(w,s)\tilde{g}_{1}(w,s)} \right) ||  \tilde{g}_{1} - \tilde{g}_{n,1} ||_{P_X} || \bar{Q}_{Y} - \bar{Q}_{n,Y} ||_{P_X} \\
    &\hspace{0.2in} = o_{\text{p}}(n^{-1/2}) \ ,
\end{align*}
which follows from Assumptions A3 and A4. Likewise, we can bound $R_{2,3}^F(\hat P_X, P_X)$, 
\[R_{2,3}^F(\hat{P}_X, P_X)\leq \left\{\sup_{w,s}\frac{\tilde{g}_{n,2}(w,s)g_{n,C,1}(w)}{\tilde{g}_{n,1}(w,s)}\right\}||g_{C,1} -g_{n,C,1}||\,||\bar{Q}_{Y} - \bar{Q}_{n,Y}||=o_{\text{p}}(n^{-1/2}) \ , \]
as well as $R_{2}$, \begin{align*}
    R_2(\hat{P}_X, P_X, \hat{P}, P) &\le \left\{\sup_{v} g_{n,R}^{-1}(1 \mid v)\right\} || g_{n,R}(1 \mid \cdot) - g_{R}(1 \mid \cdot) ||_P || \bar{Q}_{n, D(P_X)} - \bar{Q}_{D(\hat{P}_X)} ||_P \\
    &= o_{\text{p}}(n^{-1/2}) \ . 
\end{align*}
Thus, $\psi_{n,1}^+(a_1, a_2) - \psi(a_1, a_2) = P_n D(P) + o_{\text{p}}(n^{-1/2})$. Because $O_1, \dots, O_n$ are independent, the central limit then implies the result.

The proof of Theorem 2 follows almost exactly as above. We start by noting that based on assumption B7, we have \[
\psi_{n,2}^+(a_1, a_2) - \psi(a_1, a_2) = P_n D(P) + R_2(\hat{P}_X, P_X, \hat{P}, P) + o_{\text{p}}(n^{-1/2}) \ . 
\]
All that remains is to establish the negligibility of the remainder terms as written in Corollary 1. This proceeds similarly as above and we can establish that under the conditions of the theorem \begin{align*}
    R_{2,1}^F(\hat{P}_X', P_X) &= P_X \left\{ g_{n,2}^{-1} \  (\tilde{Q}'_{n,\tilde{Q}(W,A,C,CY)} - \tilde{Q}_{\tilde{Q}(W,A,C,CY)}) \ (g_{n,2} - g_2) \right\} \\
    &\le \left\{\underset{w}{\mbox{sup}} \  g_{n,2}^{-1}(w) \right\} || \tilde{Q}'_{n,\tilde{Q}(W,A,C,CY)} - \tilde{Q}_{\tilde{Q}(W,A,C,CY)} ||_{P_X} ||g_{n,2} - g_2 ||_{P_X} \\ 
    &= o_{\text{p}}(n^{-1/2}) \ , 
\end{align*}
as well as that \begin{align*}
    R_{2,1}(\hat{P}_X, P_X, \hat{P}, P) &= P \left[ g_{n,R}^{-1}(1 \mid \cdot) \{g_R(1 \mid \cdot) - g_{n,R}(1 \mid \cdot)\} \{\tilde{Q}_{D(\hat{P}_X)} - \tilde{Q}'_{n, D(P_X)} \}\right] \\
    &\le \left\{\underset{v}{\mbox{sup}} \ g_{n,R}^{-1}(1 \mid v) \right\} ||g_R(1 \mid \cdot) - g_{n,R}(1 \mid \cdot) ||_P || \tilde{Q}_{D(\hat{P}_X)} - \tilde{Q}'_{n, D(P_X)} ||_P \\
    &= o_{\text{p}}(n^{-1/2}) \ , 
\end{align*}
and
\begin{align*}
    R_{2,2}(\hat{P}_X, P_X, \hat{P}, P) &\le \left[\underset{v}{\mbox{sup}} \ \{g_{n,R}^2(1 \mid v) \ g_{n,2}(w) \}^{-1} \right] || g_R(1 \mid \cdot) - g_{n,R}(1 \mid \cdot) ||_P \\
    &\hspace{2.4in} \times ||\tilde{Q}_{\bar{Q}_{n,Y}(W,S)} - \tilde{Q}_{n,\bar{Q}_Y(W,S)} ||_P \\
    &= o_{\text{p}}(n^{-1/2}) \ .
\end{align*}

\section{B. Theoretical comparison of the classic and alternative one-step approaches}

The main difference between the classic and alternative approach is that the latter does not involve inverse probability weights when integrating $\bar{Q}_Y(W,S)$ over $S$. Instead, we first integrate out $S$ over the conditional distribution of $S$ given $R = 1, W, A, C, CY$. By assumption, this conditional distribution is is equal to the conditional distribution of $S$ given $W, A, C, CY$ in the full data distribution, so there is no need to bother with inverse probability weights in its estimation. A second regression is then applied to integrate out $A, C,$ and $CY$ to obtain the appropriate function of $W$. This strategy may have a theoretical advantage in situations where $g_R$ is unknown. To understand why, note that the multiple robustness of the classic approach allows for inconsistent estimation of $g_R$, so long as $\bar{Q}_{n,D}$ is consistent. However, estimation of the sequential regressions $\bar{Q}_Y$ and $\bar{Q}_{\bar{Q}_Y}$ relies on inverse weighting by the estimate of $g_R$. Thus, if $g_R$ is inconsistently estimated, it is likely that these quantities will be inconsistently estimated as well. In this case, consistency of the estimator is fully reliant on consistent estimation of the other nuisance quantities. On the other hand, the alternative one-step estimator decreases the reliance of the estimator on $g_R$ by replacing one step of inverse weighted estimation. Thus, we hypothesize that the alternative one-step may enjoy better performance in situations where $g_R$ is inconsistently estimated. However, this is not fully satisfactory, as estimation of the sequential regression quantity $\tilde{Q}_{\bar{Q}_Y}$ still relies on estimation of $\bar{Q}_Y$, which itself relies on estimation of $g_R$. Thus, the alternative estimator may also be subject to poor performance when $g_R$ is inconsistently estimated. 

\subsection{C. A one-step estimator that does not rely on inverse-weighted learning}

In fact, we can outline an estimation approach that fully disentangles estimation of $g_R$ and estimation of the required full data nuisance parameters. However, the approach does not appear to scale well in settings where $S$ is continuous valued. 

Let $q_S(s \mid W, A, Y)$ be defined as the conditional density of $S$ given $R = 1, W, A, C = 1, Y$ evaluated at $s$. Note that this parameter of $P$ is the same as the conditional density of $S$ given $W, A, C = 1, Y$ under $P_X$. For each $s \in \mathcal{S}$, we define the nuisance quantities $\bar{Q}_{Y q_S}(s \mid W) := E\{Y q_S(s \mid W, A = a_1, C = 1, Y) \mid A = a_1, C = 1, W \}$ and $\bar{Q}_{q_S}(s \mid W) := E\{ q_S(s \mid W, A = a_1, Y) \mid A = a_1, W \}$. We now derive a relationship between $E_X(CY \mid C = 1, W, A = a_1, S)$ and these two quantities. For simplicity and without loss of generality we show the derivation when data are discrete. \begin{align*}
&E_X(CY \mid A = a_1, C = 1, W, S = s) \\
 &\hspace{0.2in} = \sum_y y \ P_X(CY=y\mid C = 1, S=s,A=a,W)\\
&\hspace{0.2in} =\sum_y y \ \frac{P_X(S=s\mid C = 1, Y = y, W, A=a)}{P_X(S=s\mid C = 1, W, A=a_1)}P_X(CY=y\mid C = 1, A=a_1,W=w)\\
&\hspace{0.2in} =\sum_y \left\{ y \ \frac{P_X(S=s\mid C = 1, Y=y, W, A=a_1) P_X(Y=y\mid C = 1, A=a_1,W)}{\sum_y P_X(S=s\mid C = 1, Y=y, W, A=a_1)P_X(Y=y\mid C = 1, A=a_1,W)}\right\}\\
&\hspace{0.2in} =\sum_y \left\{ y \ \frac{P(S=s\mid C = 1, Y=y, W, A=a_1, R=1) P(Y=y\mid C = 1, A=a_1,W)}{\sum_y P(S=s\mid C = 1, Y=y, W, A=a_1, R=1)P(Y=y\mid C = 1, A=a_1,W)} \right\}\\
&\hspace{0.2in} = \frac{\sum_y y \ P(S=s\mid C = 1, Y=y, W, A=a_1, R=1) P(Y=y\mid C =1, A=a_1,W)}{\sum_y P(S=s\mid C = 1, Y=y, W, A=a_1, R=1)P(Y=y\mid C =1, A=a_1,W)} \\
&\hspace{0.2in} = \frac{\bar{Q}_{Y q_S}(s \mid W)}{\bar{Q}_{q_S}(s \mid W)}
\end{align*}
Thus, we have the following identification result, \[
\psi(a_1, a_2) =  E\left( E\left[ E\left\{ \frac{\bar{Q}_{Y q_S}(S \mid W)}{\bar{Q}_{q_S}(S \mid W)} \mid R = 1, W, A, C, CY \right\} \mid A = a_2, W \right] \right) \ . 
\]
This suggests the following estimation strategy for evaluating the plug-in estimator. First, obtain an estimate $q_{n,S}$ of the conditional mediator density $q_S$. For all $i$ such that $R_i = 1$, evaluate $q_{n,S}(S_i \mid W = W_j, A = a_1, Y = Y_j)$ for all $j$ such that $C_j = 1$ and $A_j = a_1$. Next, obtain an estimate $\bar{Q}_{n, Y q_S}(S_i \mid W)$ of $\bar{Q}_{Y q_S}(S_i \mid W)$ by using observations such that $C_j = 1$ and $A_j = a_1$ to regress the outcome $Y_j \times q_{n,S}(S_i \mid W = W_j, A = a_1, Y = Y_j)$ onto $W_j$. Similarly, obtain an estimate $\bar{Q}_{n,q_S}(S_i \mid W)$ of $\bar{Q}_{q_S}(S_i \mid W)$ by regressing the outcome $q_{n,S}(S_i \mid W = W_j, A = a_1, Y = Y_j)$ onto $W_j$. Next, for all $i$ such that $R_i = 1$, evaluate $\bar{Q}_{n, Y q_S}(S_i \mid W_i) / \bar{Q}_{n,q_S}(S_i \mid W_i)$. This quantity serves as our estimate of $\bar{Q}_{n,Y}$ and we proceed as with the plug-in estimator for the alternative one-step. 

The key element of this estimator's construction is that it does not rely on inverse weighting by an estimate of $g_R$ to estimate nuisance parameters. However, there is a price to be paid in terms of complexity. Whereas the previous estimators each allowed us to avoid estimation of the conditional mediator density, in this approach we require such an estimator. Moreover, we require a separate estimation of $\bar{Q}_{q_S}$ and $\bar{Q}_{q_S}$ {\it for each value} at each value $S_i$ such that $R_i = 1$. Thus, if $S$ is continuous-valued, this would require fitting $2\sum_{i=1}^n R_i$ regressions. This strategy therefore appears most-appealing in situations where $S$ is binary, in which case its implementation is quite tractable, both in terms of density estimation and in terms of the estimation 
of $\bar{Q}_{q_S}$ and $\bar{Q}_{q_S}$.

\section*{D. Additional simulation results}

We found that the estimates based on a misspecified GLM performed quite poorly in terms of their bias (Table \ref{sim_with_glmmain}). Not only that, the variance of the estimators was extremely inflated and confidence intervals tended to have conservative coverage probabilities due to standard error estimates that were too large. 

Table \ref{risk_est_table} shows estimators of the counterfactual risk $\psi(1,0)$. We found similar results as with the estimators of indirect effect and proportion mediated. Confidence intervals attained near nominal coverage in larger samples with negligible bias. 

\begin{table}[!htbp]
\centering
\caption{Comparison of estimators of indirect effect and proportion mediated for the COVID-19 simulation including a misspecified main terms generalized linear model.}
\label{sim_with_glmmain}
\resizebox{\columnwidth}{!}{%
\begin{tabular}{rrrrrrrrrl}
  & \multicolumn{4}{c}{$\psi_{n,1}^{+}$} & \multicolumn{4}{c}{$\psi_{n,2}^{+}$} & \\ 
 \hline
  & \multicolumn{2}{c}{Indirect} & \multicolumn{2}{c}{Prop. Med.} & \multicolumn{2}{c}{Indirect} & \multicolumn{2}{c}{Prop. Med.} & \\ 
 $\alpha$ & Bias & Coverage & Bias & Coverage & Bias & Coverage & Bias & Coverage & Method \\ 
 \hline
-5.0 & 3.563 & 0.998 & -0.276 & 0.999 & 3.526 & 0.844 & -0.255 & 0.856 & \multirow{5}{6em}{GLM main} \\ 
  -4.1 & -1.762 & 0.995 & -0.047 & 0.994 & 1.969 & 0.952 & -0.100 & 0.951 &  \\ 
  -3.6 & 1.521 & 0.996 & 0.040 & 0.995 & 0.477 & 0.972 & -0.043 & 0.974 & \\
  -3.3 & -0.253 & 0.999 & 0.099 & 0.999 & 0.109 & 0.979 & -0.022 & 0.980 & \\
  -3.1 & -0.303 & 0.999 & 0.110 & 0.999 & 0.085 & 0.982 & -0.013 & 0.982 & \\
   \hline
-5.0 & 4.361 & 0.842 & -0.270 & 0.857 & 3.509 & 0.805 & -0.256 & 0.823 & \multirow{5}{6em}{GLM inter.} \\ 
  -4.1 & 1.103 & 0.922 & -0.098 & 0.926 & 1.807 & 0.911 & -0.100 & 0.916 &  \\ 
  -3.6 & 0.828 & 0.935 & -0.048 & 0.938 & 0.390 & 0.930 & -0.042 & 0.933 & \\
  -3.3 & 0.075 & 0.947 & -0.023 & 0.947 & 0.850 & 0.943 & -0.024 & 0.948 & \\
  -3.1 & 0.025 & 0.945 & -0.013 & 0.950 & 0.083 & 0.943 & -0.013 & 0.948 & \\
   \hline
-5.0 & -0.843 & 0.903 & -0.202 & 0.911 & 1.528 & 0.876 & -0.192 & 0.889 & \multirow{5}{6em}{SuperLearner} \\ 
  -4.1 & 0.396 & 0.947 & -0.079 & 0.947 & 0.375 & 0.927 & -0.065 & 0.931 & \\ 
  -3.6 & 0.261 & 0.977 & -0.052 & 0.979 & 0.187 & 0.959 & -0.035 & 0.960 & \\
  -3.3 & -0.677 & 0.981 & -0.039 & 0.982 & 0.139 & 0.963 & -0.030 & 0.967 & \\
  -3.1 & 0.069 & 0.982 & -0.030 & 0.983 & 0.122 & 0.970 & -0.027 & 0.971 & \\
  \hline
\end{tabular}%
}
\end{table}

\begin{table}[ht]
\centering
\caption{Performance of the two one-step estimators of $\psi(1,0)$ in terms of bias, coverage probability of a 95\% confidence interval, and the median estimated standard error divided by the true standard error.}
\label{risk_est_table}
\resizebox{\columnwidth}{!}{%
\begin{tabular}{rrrrrrrrl}
  & &\multicolumn{3}{c}{$\psi_{n,1}^{+}(1,0)$} & \multicolumn{3}{c}{$\psi_{n,2}^{+}(1,0)$} & \\ 
 \hline
  $\alpha$ & $\psi(1,0)$ & Bias & Coverage & $\frac{\text{Est. std.}}{\text{True. std.}}$ & Bias & Coverage & $\frac{\text{Est. std.}}{\text{True. std.}}$ & Method \\ 
 \hline
-5.0 & 0.0016 & 0.0002 & 0.6250 & 5.0116 & 0.0001 & 0.6220 & 0.8528 & \multirow{5}{6em}{GLM main} \\ 
  -4.1 & 0.0038 & 0.0003 & 0.8990 & 4.8306 & 0.0002 & 0.8730 & 1.2171 &  \\ 
  -3.6 & 0.0063 & -0.0002 & 0.9790 & 4.3200 & 0.0004 & 0.9410 & 1.2139 & \\
  -3.3 & 0.0085 & 0.0002 & 0.9960 & 4.0146 & 0.0005 & 0.9510 & 1.2209 & \\
  -3.1 & 0.0103 & 0.0002 & 0.9970 & 3.8029 & 0.0007 & 0.9640 & 1.2445 & \\ 
   \hline
-5.0 & 0.0016 & 0.0001 & 0.6220 & 0.6468 & 0.0001 & 0.6210 & 0.6454 & \multirow{5}{6em}{GLM inter.} \\ 
  -4.1 & 0.0038 & 0.0002 & 0.8380 & 0.9314 & 0.0002 & 0.8320 & 0.9293 &  \\ 
  -3.6 & 0.0063 & 0.0003 & 0.8920 & 0.9224 & 0.0003 & 0.8920 & 0.9130 & \\ 
  -3.3 & 0.0085 & 0.0005 & 0.9160 & 0.9543 & 0.0005 & 0.9140 & 0.9418 & \\ 
  -3.1 & 0.0103 & 0.0007 & 0.9250 & 0.9649 & 0.0007 & 0.9230 & 0.9551 & \\ 
  \hline
-5.0 & 0.0016 & 0.0001 & 0.6190 & 0.6909 & 0.0001 & 0.6170 & 0.8796 & \multirow{5}{6em}{SuperLearner} \\ 
  -4.1 & 0.0038 & 0.0002 & 0.8480 & 1.0329 & 0.0002 & 0.8760 & 1.0659 & \\ 
  -3.6 & 0.0063 & 0.0003 & 0.9000 & 1.0239 & 0.0002 & 0.9380 & 1.0530 & \\ 
  -3.3 & 0.0085 & 0.0003 & 0.9200 & 1.0884 & 0.0004 & 0.9610 & 1.0687 & \\ 
  -3.1 & 0.0103 & 0.0004 & 0.9370 & 1.1188 & 0.0004 & 0.9720 & 0.9704 & \\ 
   \hline
\end{tabular}%
}
\end{table}

\begin{table}[ht]
\centering
\caption{Proportion of estimates of $\psi(1,0)$ that were less than zero for the classic one-step estimator}
\begin{tabular}{rrl}
  \hline
  $\alpha$ & Prop. $\psi_{n,1}^{+}(1,0)<0$ & Method \\ 
 \hline
-5.0 & 0.10 & \multirow{5}{6em}{GLM main} \\ 
  -4.1 & 0.21 &  \\ 
  -3.6 & 0.25 &  \\ 
  -3.3 & 0.28 &  \\ 
  -3.1 & 0.28 &  \\ 
   \hline
-5.0 & 0.01 & \multirow{5}{6em}{GLM inter.} \\ 
  -4.1 & 0.00 &  \\ 
  -3.6 & 0.00 &  \\ 
  -3.3 & 0.00 &  \\ 
  -3.1 & 0.00 &  \\ 
   \hline
-5.0 & 0.01 & \multirow{5}{6em}{SuperLearner} \\ 
  -4.1 & 0.03 &  \\ 
  -3.6 & 0.04 &  \\ 
  -3.3 & 0.04 &  \\ 
  -3.1 & 0.05 &  \\ 
   \hline
\end{tabular}
\end{table}

\end{document}